\documentclass[twoside,journal]{IEEEtran}

\usepackage{cite,graphicx,amssymb,amsmath,color,textcomp,tabularx}
\usepackage[rgb]{xcolor}

\usepackage[caption=false,font=footnotesize]{subfig}
\usepackage{siunitx} 
\usepackage{mathrsfs}

\usepackage{booktabs}  
\usepackage{threeparttable}

\IEEEoverridecommandlockouts

\renewcommand{\arraystretch}{0.78}

\def \Tr {\mathrm{Tr}}
\usepackage{graphicx}
\usepackage{epstopdf}

\usepackage{algorithm,algorithmic}
\usepackage{setspace}

\usepackage{bm}
\usepackage{multirow}

\IEEEoverridecommandlockouts
\begin{document}

\title{Rethinking Secure Precoding via Interference Exploitation:   A Smart Eavesdropper  Perspective
}
\author{
\IEEEauthorblockN{Qian Xu, \emph{Student Member, IEEE},  Pinyi Ren,  \emph{Member, IEEE},\\ and A. Lee Swindlehurst, \emph{Fellow, IEEE} }
\thanks{Qian Xu and Pinyi Ren are with the School of Electronic and Information Engineering,
	Xi'an Jiaotong University, Xi'an, China  (e-mails: xq1216@stu.xjtu.edu.cn; pyren@mail.xjtu.edu.cn).
	
	A. Lee Swindlehurst is with  the Center for Pervasive
	Communications and Computing,  University of
	California, Irvine, CA 92697, USA  (e-mail: swindle@uci.edu).

}

}

\maketitle

\begin{abstract}
Based on the concept of constructive interference (CI), multiuser interference (MUI) has recently been shown to be beneficial for   communication secrecy.  A few CI-based secure precoding algorithms have been proposed that use both the channel state information (CSI) and knowledge of the instantaneous transmit symbols.  In this paper, we examine the  CI-based secure precoding problem with a focus on smart eavesdroppers that exploit statistical information gleaned from the precoded data for symbol detection. Moreover, the impact of correlation between the main and eavesdropper channels is taken into account. We first modify an existing CI-based preocding scheme to better utilize the destructive impact of the interference. Then,  we point out the drawback of both  the existing and the new modified CI-based precoders when faced with a smart eavesdropper. To address this deficiency, we provide a general  principle for precoder design and then give  two specific  design examples. Finally, the scenario where the eavesdropper's CSI is unavailable is  studied. Numerical results show that although our modified CI-based  precoder can achieve a better  energy-secrecy trade-off than the existing approach, both have a limited secrecy benefit. On the contrary, the precoders developed using the new CI-design principle can achieve a much improved trade-off and significantly degrade the eavesdropper's  performance.

\end{abstract}


%\ls{0.9}

\section{Introduction}\label{sec_introduction}

Multiuser interference (MUI) is usually considered  harmful  for  downlink multiuser communication systems. The MUI can greatly limit the achievable rate of each user especially for a dense cellular network with many users. One common approach to address MUI is multiuser precoding~\cite{Spencer,Sadek,Christensen,SchubertBoche2004,Sidiro}, which typically employs a linear precoder at the transmitter to suppress or completely remove the MUI at each  user's receiver. Zero-forcing is a common technique to eliminate MUI~\cite{Spencer}, but it can lead to noise amplification in certain scenarios. As an alternative, signal-to-leakage-and-noise ratio can be chosen as the design metric~\cite{Sadek}, which leads to a closed-form result. For other approaches,  the precoders are  generally obtained by solving one or two kinds of optimization problems: maximizing the sum rate under a total power constraint (e.g.,~\cite{Christensen}) or minimizing transmit power under a quality of service (QoS) requirement at each user (e.g.,~\cite{SchubertBoche2004,Sidiro}).

Instead of suppressing MUI,  recent research~\cite{MasourosRat} suggests that known interference can be exploited as a  useful source for improving the received signal power. According to~\cite{MasourosRat}, the interference is regarded as  constructive interference (CI) if it contributes to a  power enhancement of the desired symbol; otherwise, it is  destructive. By exploiting the information about the symbols to be transmitted and the  channel state,  the MUI can be designed in some cases to be constructive for the desired users. Following this idea, an increasing number of studies on  CI-based precoding have been proposed~\cite{Masouros2,LiMasouros2018,Haqiqat,HJedda,Ang2}, where~\cite{Masouros2,LiMasouros2018,Haqiqat} focused on  regular multiple-input multiple-output (MIMO) systems,  while~\cite{HJedda,Ang2} considered  massive MIMO systems with low-resolution digital-to-analog converters.  These CI-based  approaches can reduce the required transmit power to achieve a certain level of performance, especially when the system is heavily loaded. However, they are mainly useful for simple constellation designs like phase-shift keying
(PSK), and they require symbol-level precoding, which is more complex.

In recent years, there has been a growing interest in improving the security of wireless communications systems. Physical layer security (PLS)~\cite{Wyner} is a technique that exploits the  randomness of wireless channels to achieve secure transmission. Through an appropriate design of the transmit signal, PLS techniques can be used to minimize the power of the eavesdropper's received signal while guaranteeing a certain detection performance at the desired user~\cite{Goel,Wei,J.Zhuu,ZhiDing}. CI-based precoding already benefits secrecy since it can achieve the same performance for a given link with reduced transmit power, which already reduces an eavesdropper's ability to decode the sensitive data.  There has been some prior work that employs a CI-based approach to 
improve communication secrecy~\cite{kat1,kat2,Ont}. In~\cite{kat1,kat2}, CI-based  secure precoding algorithms were proposed to achieve energy-efficient secure message transmission and secure simultaneous wireless
information and power transfer (SWIPT), respectively, assuming  single-antenna eavesdroppers. Specifically, with the channel state information (CSI) available of all users' channels, a constructive-destructive (C-D) interference-based secure precoding method\footnote{This precoding method is referred to as C-D precoding algorithm in the following paragraphs. The details of this precoder will be given in Section~\ref{sec_cons_des_precoding_task}.}  was proposed, which exploits the MUI to push the legitimate user's received signal into the constructive region for the desired symbol, and push the eavesdropper's noise-free received signal outside the constructive region. The authors in~\cite{Ont}  studied 
a related problem with a multi-antenna eavesdropper that can obtain a better estimate of the transmitted symbol. Without the eavesdropper's CSI,  several power-minimizing  precoding  algorithms were proposed.

Similar to~\cite{kat1,kat2}, in this paper we study a single-antenna eavesdropper. We focus on the detection performance at the eavesdropper rather than the secrecy rate, since CI-based precoding assumes simple fixed signal constellations such as PSK or quadrature amplitude modulation (QAM), which is related to the prior work on secure transmission with finite constellations~\cite{ZhiDing}. 
It is important to note that the eavesdropper in~\cite{kat1,kat2}  is assumed to adopt the same  detection method as the legitimate user. For that simple detector, the detection rule is simply to choose the constellation point  nearest  the received signal. For this simple eavesdropper, the  C-D  precoding method in~\cite{kat1,kat2} can yield very good secrecy performance  since  the eavesdropper's received signal  is designed to be far away from the real transmitted symbol. 
However, if the eavesdropper can exploit the  statistical characteristics  of  the received signal, a much better detection performance can be achieved. For example,  for  PSK modulation, the eavesdropper can exploit knowledge of the constellation used, the QoS parameters and the channel distribution to obtain a (possibly empirical) model for the conditional distribution of the received signal phase $\theta_e$ given the transmitted symbols. This information can then in turn be used to enable the eavesdropper to implement an optimal maximum likelihood (ML) detector. 
In this paper, an eavesdropper that can learn and implement this ML detector is referred to as a {\emph{smart}} eavesdropper in this paper.
When faced with a smart eavesdropper, the  C-D precoding algorithm may unexpectedly reveal the transmitted symbols. since the  restriction on the location of the eavesdropper's received signal makes the distribution of $\theta_e$ non-uniform on $[0,2\pi]$, which helps the eavesdropper  distinguish  different   symbols. 

To address the above issue, in this paper we restudy  CI-based secure precoding from the perspective of a smart eavesdropper. Furthermore, we consider a general scenario where the eavesdropper channel is possibly correlated with the main channel, which makes it easier for the eavesdropper to observe the desired symbol. We first review the C-D precoding algorithm in~\cite{kat1,kat2}. It is observed that although the eavesdropper's noise-free received signal is kept outside the constructive region, the scheme in~\cite{kat1,kat2} only exploits a part of the destructive region, which significantly increases the transmit power. Inspired by this observation, we propose our own C-D precoding algorithm, which exploits the {\em full} destructive region and thus  can save power. Then, we  point out the drawback of  the above C-D precoding strategies in the presence of a smart eavesdropper. To overcome  the drawback, we propose another secure precoding method, which  only  limits the power  of the eavesdropper's received signal  without any constraints on the phase. Following this, a low-complexity secure precoding algorithm is also proposed. Finally, we study a more practical scenario where the eavesdropper's CSI is unavailable. The primary contributions of the paper are enumerated below.
\begin{enumerate}
	\item We show that the C-D approach of~\cite{kat1,kat2} only exploits a portion of the destructive region when designing the precoder.
	\item We present a modification of the C-D approach of~\cite{kat1,kat2} that exploits the full destructive region, and that is thus able to achieve the same level of security with less transmit power.
	\item We show that, for PSK signals, the C-D precoding approach is susceptible to a smart eavesdropper that can derive the conditional distribution of the phase given the transmitted signals.
	\item We present a new design principle for secure CI-based precoding that addresses this deficiency, and we propose two algorithms based on this principle to generate the precoder. While these new approaches will require an increase in transmit power compared with C-D precoding, they can achieve significantly improved security that approaches the best possible performance.
	\item Unlike the methods above which assume availability of the eavesdropper's CSI, we develop an alternative algorithm that uses artificial noise (AN) for the case where the eavesdropper's CSI is unavailable.
\end{enumerate}

The rest of this paper is organized as follows. Section~\ref{sec_system_model} gives the system model and the basic idea for constructive interference. Section~\ref{sec_traditional_cons_des}  first reviews the existing C-D precoding algorithm  and then presents an improved version of the C-D precoding algorithm. After introducing the smart eavesdropping approach, the security risk of the C-D precoding methods is  revealed. Section~\ref{sec_secure_precoding_smart_eavesdropper}  gives a general principle for designing the precoder against  a smart eavesdropper. Then, two specific precoding algorithms are provided. Section~\ref{sec_noCSI} studies the scenario where the eavesdropper's CSI is unavailable. Section~\ref{sec_simulation_results} presents numerical results to evaluate  the performance  of all the  precoding schemes, and finally  Section~\ref{sec_conclusions} concludes the paper.

\emph{Notations:} In this paper,  bold upper-case letters denote matrices while bold  lower-case letters  denote vectors. The set of $M \times N$ complex matrices is denoted as  $\mathbb{C}^{M \times N}$. For a vector $\mathbf{x}$,  $\mathbf{x}^T$, $\mathbf{x}^H$,  and $\|\mathbf{x}\|$ represent the  transpose, Hermitian transpose, and  $l_2$ norm of $\mathbf{x}$, respectively. For a scalar $x$, 
$|x|$  and $x^*$ represent the $l_2$ norm and complex conjugate of $x$, respectively.   $\mathbf{I}$ denotes the identity matrix  while   $\mathbf{1}_{N\times 1}$ represents an $N\times 1$ vector composed of all ones.  Finally, 
$ \mathcal{CN}(\mathbf{\bm{\mu}},\mathbf{\Omega})$ denotes the complex circular Gaussian distribution with mean $\mathbf{\bm{\mu}}$ and covariance $\mathbf{\Omega}$.

\section{System Model} \label{sec_system_model}

\begin{figure}
	\centering
	\includegraphics[width=2.3in]{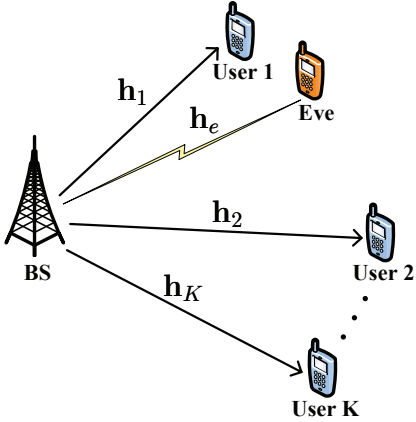}
	\caption{ System model: A downlink multiuser communication network with an eavesdropper attempting to get user 1's message.}
	\label{system_model}
\end{figure}

\subsection{System Description}\label{sec_system_description}
As illustrated in Fig.~\ref{system_model}, we concentrate on  downlink transmission in  a  single-cell multiuser system. The base station (BS), equipped with $N$ antennas,  simultaneously serves $K$  single-antenna users  using the same time-frequency resource. Apart from the  users being served, there  also exist  idle users who do not have permission to access the  transmitted information but may try to overhear it.  These idle users are potential eavesdroppers. In this paper, we consider a simple but typical scenario where  a single eavesdropper is located near one of the users (labeled as user 1)  to try to wiretap the  confidential message for this user. The other users are assumed to be ordinary users without a secure communication requirement. Like the $K$ users,  the  eavesdropper is also  equipped with a single antenna. 
%Therefore, specific transmit mechanisms are needed to obstruct the  eavesdropping.

At the BS, each element of the symbol vector $\mathbf{s}=[s_1, \ldots, s_K]^T$,  where  $s_k$ is the desired symbol for user $k$, is drawn from a normalized \emph{M}-PSK constellation set $\mathcal{C}_M=\left\{c_m:m=1,\cdots,M\right\}$ given by
\begin{equation}\label{equation-1}
c_m=\exp\left(j(2m-1)\Phi\right),
\end{equation} 
where $\Phi=\frac{\pi}{M}$. In addition, we assume that the BS has  perfect CSI of  the legitimate channel  $\mathbf{H}=\left[ \mathbf{h}_1,\cdots,\mathbf{h}_K \right]^T$, where $\mathbf{h}_k\in\mathbb{C}^{N\times 1}$ denotes the channel from the BS to user $k$. As for the eavesdropper channel $\mathbf{h}_e\in\mathbb{C}^{N\times1}$, we first assume that  full knowledge of $\mathbf{h}_e$ is available. The case where information about $\mathbf{h}_e$ is unavailable  will be studied  in Section~\ref{sec_noCSI}.  The BS  designs the  transmit signal vector $\mathbf{x}=\mathcal{P}\left(\mathbf{s},\mathbf{H},\mathbf{h}_e\right)\in\mathbb{C}^{N\times1}$ based on the symbol vector and  CSI, where  the function   
 $\mathcal{P}\{\cdot\}$  represents  a general symbol-wise precoder which can be linear or non-linear.   

The received signal at user $k$ can be expressed as
\begin{equation}\label{equation-2}
y_k=\mathbf{h}_k^T\mathbf{x}+n_k,
\end{equation} 
where $n_k\sim \mathcal{CN}(0,1)$ is  additive white Gaussian noise  at the receiver. For ease of notation, we assume that all the nodes have the same noise power, which is  normalized to be 1. In addition, we  only consider flat fading  channels and assume that $\mathbf{h}_k \in  \mathcal{CN}(\mathbf{0},\beta_k\mathbf{I})$, where $\beta_k$ models large-scale fading such as geometric attenuation and shadowing.  Similarly, the received signal at the eavesdropper is given by
\begin{equation}\label{equation-3}
y_e=\mathbf{h}_e^T\mathbf{x}+n_e,
\end{equation} 
where $\mathbf{h}_e \in  \mathcal{CN}(\mathbf{0},\beta_e\mathbf{I})$ with $\beta_e$  denoting the large-scale fading parameter and $n_e\sim \mathcal{CN}(0,1)$. Since the eavesdropper is located very close to user 1, the two channels $\mathbf{h}_1$ and $\mathbf{h}_e$ are in general correlated. Using the channel correlation model in~\cite{datong,CL}, the eavesdropper channel can be written as
\begin{equation}\label{equation-4}
\mathbf{h}_e=\sqrt{\beta_e}\left(\sqrt{\rho}\tilde{\mathbf{h}}_1+\sqrt{1-\rho}\mathbf{w} \right),
\end{equation} 
where  $\tilde{\mathbf{h}}_1=\mathbf{h}_1/\sqrt{\beta_1}$ is the normalized version of $\mathbf{h}_1$ with respect to the large-scale fading parameter,  and $\mathbf{w}  \in  \mathcal{CN}(\mathbf{0},\mathbf{I})$ is a random vector independent of $\tilde{\mathbf{h}}_1$.  The parameter
$\rho\in[0,1]$ measures the strength of the channel correlation.    Note that since the eavesdropper is close to user 1,  it will usually be true that $\beta_e=\beta_1$.    

%Generally, for symbol detection at the $K$ legitimate users,  each user needs to rescale the received signal $y_k$ first by a real-valued coefficient~\cite{HJedda}.  Otherwise, the received signal would not lie in the  decision region of the  desired symbol due to  multi-antenna gain. After the rescaling operation,  symbol detection is performed. The detector selects the constellation point that has the shortest distance to the  received signal as the detection result.  It is worth noting that for PSK symbols, there is no need to rescale the received signal  since the  transmitted symbols are only distinguished by phase. Therefore, in the following analysis the rescaling operation is omitted for  symbol detection at both legitimate users and the eavesdropper. 

The existing research in~\cite{kat1,kat2} assumes that the eavesdropper adopts the same detection method as the legitimate users, which greatly limits the eavesdropper's capability. In this paper, we consider a smart eavesdropper that can use  statistical information to improve detection performance. The details will be given in Section~\ref{sec_compromise_cons_des}. 

\subsection{Constructive Interference}\label{sec_constructive_interference}

\begin{figure}
	\centering
	\includegraphics[width=3.2in]{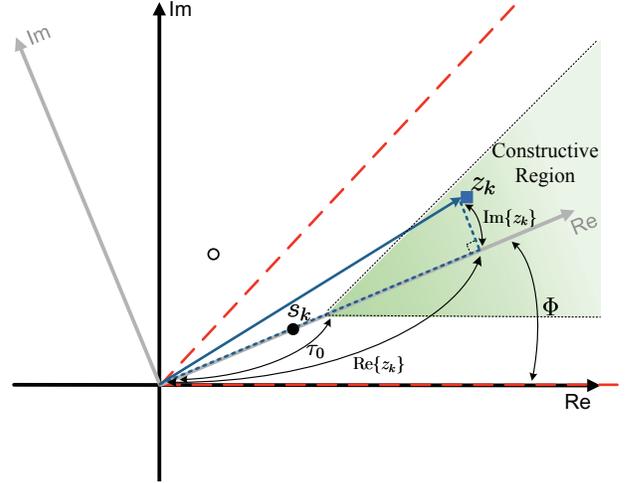}
	\caption{ Illustration of the constructive interference design for 8PSK modulation. The solid circle $s_k$ is the symbol of interest while the hollow circle is the adjacent constellation point.}
	\label{constructive_PSK}
\end{figure}

Constructive interference-based precoding transforms the undesirable MUI into useful power to push the received signal further away from the \emph{M}-PSK decision  boundaries, which in turn  reduces the symbol error rate at the end user. 
To show the  concept of constructive interference  more clearly, we give an  intuitive example in Fig.~\ref{constructive_PSK},   where the symbol of interest $s_k$ is assumed to be one of the constellation points of the 8PSK  constellation. As illustrated in Fig.~\ref{constructive_PSK},  the constructive region for $s_k$  is the green   sector with infinite radius and angle $2\Phi$, and the decision region for $s_k$ is the  sector determined by the two dashed red rays which are also referred to as  decision boundaries. The distance between  the constructive region and the decision boundary  depends on  $\tau_0$, which is also related to the minimum signal-to-noise ratio (SNR) requirement. When the noise-free received signal $z_k=y_k-n_k$ lies in the constructive region, it is pushed deeper into the decision region  and thus is more robust to additive noise perturbations. With  knowledge of the CSI and the  symbols to be transmitted, CI-based precoding guarantees that the noise-free  signal received at each user lies in the  constructive region of each user's desired symbol. In this way, the  MUI is transformed into useful energy for improving the SNR.

%In this way, the MUI works as a useful  interference with double functions. On the one hand, it degrades eavesdroppers' symbol detection capabilities; on the other hand, it benefits legitimate users by improving the energy of the desired symbols. The mathematical formulations of the  signal vector design problem will be discussed as below.

\section{ Traditional  Constructive-Destructive Interference Based Secure Precoding} \label{sec_traditional_cons_des}

In this section, we first review the traditional C-D  precoding algorithm in~\cite{kat1,kat2} with eavesdropper's CSI. The idea of C-D precoding and the corresponding algorithm are described. Then, a smart eavesdropper, who can exploit statistical information for symbol detection, is introduced.  Finally, we point out the drawback of  C-D precoding when faced with a smart eavesdropper.

\subsection{Precoding Task and Solution}\label{sec_cons_des_precoding_task}

\begin{figure}[t]
	\centering
	\subfloat[]{
		\includegraphics[width=3in]{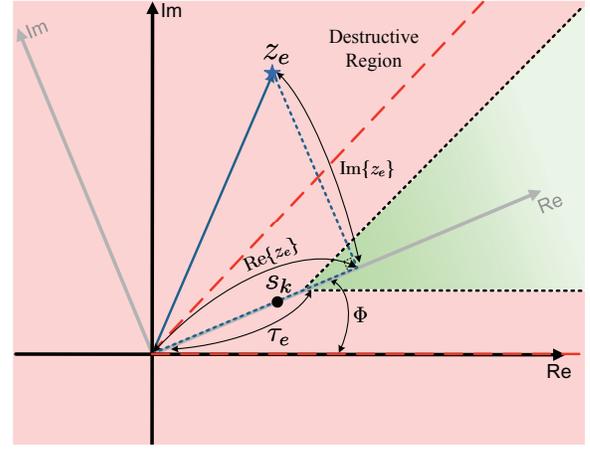}
	}
	\hfil
	\subfloat[]{
		\includegraphics[width=3in]{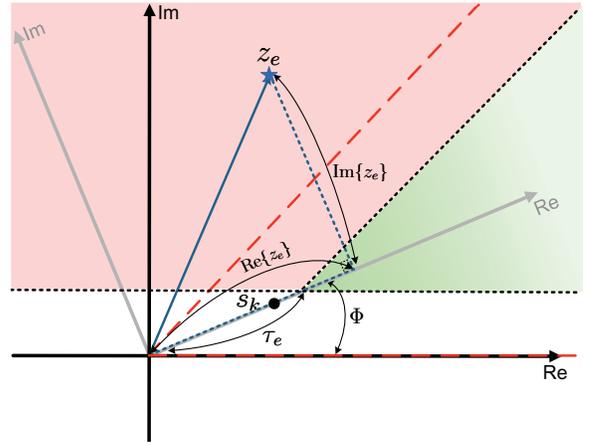}
	}
	\caption{ The constructive-destructive interference based secure precoding  for 8PSK modulation.  (a) Our proposed scheme  (b) The scheme in~\cite{kat1,kat2}.
	}
	\label{destructive_eve}
\end{figure}

As discussed in Section~\ref{sec_constructive_interference}, it is beneficial for the legitimate users if  their noise-free received signals are  located in the constructive regions of the desired symbols. On the other hand, the noise-free received signal at the eavesdropper should be located outside the constructive region, which will  make a correct detection at the eavesdropper more challenging.  Accordingly, the C-D precoding method of~\cite{kat1,kat2}  pushes the received signal at  each legitimate user towards the corresponding constructive region, while guaranteeing that the noise-free received signal at the eavesdropper lies in the destructive region (the red area in Fig.~\ref{destructive_eve}(a)). As in~\cite{kat1,kat2}, different SNR requirements are assumed for the  users and eavesdropper; in particular, it is assumed that the desired minimum SNR is $\gamma_0$ for all users, and the desired maximum SNR is $\gamma_e$ for the eavesdropper.   Generally, $\gamma_e$ is chosen to be much smaller than $\gamma_0$ to make the eavesdropper suffer a  higher symbol error rate. Since the noise power is assumed to be one, for the parameters $\tau_0$ and $\tau_e$ in Fig.~\ref{constructive_PSK} and Fig.~\ref{destructive_eve}, we have $\tau_0=\sqrt{\gamma_0}$ and $\tau_e=\sqrt{\gamma_e}$. 

Now, we formulate the  optimization problem that the C-D precoding approach of~\cite{kat1,kat2} attempts to solve.  First of all, in order to find a uniform expression for the constructive regions of different symbols in the constellation, we rotate the  original coordinate system by the phase of the  symbol of interest. Taking the symbol $s_k$ as an example, the new coordinate system after rotation is shown by the gray one in both Fig.~\ref{constructive_PSK} and  Fig.~\ref{destructive_eve}. Based on \eqref{equation-2}, in the new coordinate system, the  real and  imaginary part of the noise-free received signal $z_k$ can be respectively derived as
\begin{equation}\label{equation-5}
	\mathrm{Re}\{z_k\}=\mathrm{Re}\left\{\frac{(y_k-n_k)s_k^*}{|s_k|}\right\}=\mathrm{Re}\{\mathbf{g}_k^T\mathbf{x}\} 
\end{equation}
and
\begin{equation}\label{equation-6}
	\mathrm{Im}\{z_k\}=\mathrm{Im}\left\{\frac{(y_k-n_k)s_k^*}{|s_k|}\right\}=\mathrm{Im}\{\mathbf{g}_k^T\mathbf{x}\},
\end{equation}
where we adopt the new variable $\mathbf{g}_k\triangleq\mathbf{h}_ks_k^*$ and use the fact that $|s_k|=1$.
As illustrated in Fig.~\ref{constructive_PSK}, the constructive region for $s_k$ can  be defined by the following inequality
\begin{equation}\label{equation-7}
	\left|\mathrm{Im}\{z_k\}\right|\leq \left(\mathrm{Re}\{z_k\}-\tau_0\right)\tan \Phi,
\end{equation}
which already indicates that $\mathrm{Re}\{z_k\}\geq \tau_0$.

For convenience, we  rewrite the   constraint in \eqref{equation-7}  using  real-valued  notation. In particular, we define the following real-valued vectors 
\begin{equation}\label{equation-8-1}
	\mathbf{a}_k^T=\left[\mathrm{Re}\{\mathbf{g}_k^T\},-\mathrm{Im}\{\mathbf{g}_k^T\}\right],
\end{equation}
\begin{equation}\label{equation-8-2}
\mathbf{b}_k^T=\left[\mathrm{Im}\{\mathbf{g}_k^T\},\mathrm{Re}\{\mathbf{g}_k^T\}\right],
\end{equation}
\begin{equation}\label{equation-9}
	\bar{\mathbf{x}}^T=\left[\mathrm{Re}\{\mathbf{x}^T\},\mathrm{Im}\{\mathbf{x}^T\}\right].
\end{equation}
It can be easily verified that
\begin{equation}\label{equation-10}
	\mathrm{Re}\{z_k\}=\mathbf{a}_k^T\bar{\mathbf{x}},~\mathrm{Im}\{z_k\}=\mathbf{b}_k^T\bar{\mathbf{x}}.
\end{equation}
Therefore, the real-valued reformulation of \eqref{equation-7} can be written as
\begin{equation}\label{equation-11}
	\left|\mathbf{b}_k^T\bar{\mathbf{x}}\right|\leq \left(\mathbf{a}_k^T\bar{\mathbf{x}}-\tau_0\right)\tan \Phi,
\end{equation}
which is equivalent to the following two constraints
\begin{subequations} \label{equation-12}
	\begin{align}
		&\left(\mathbf{a}_k^T\tan \Phi-\mathbf{b}_k^T\right)\bar{\mathbf{x}}-\tau_0\tan\Phi\geq 0, \label{12-constraint1}\\
		&\left(\mathbf{a}_k^T\tan \Phi+\mathbf{b}_k^T\right)\bar{\mathbf{x}}-\tau_0\tan\Phi\geq 0  \label{12-constraint2}.
	\end{align}
\end{subequations}

According to the C-D precoding in~\cite{kat1,kat2} (e.g., refer to problem (25) in~\cite{kat1}),  the location of the eavesdropper's noise-free received signal  $z_e=\mathbf{h}_e^T\mathbf{x}$ is constrained by 
\begin{subequations} \label{equation-13}
	\begin{align}
	&-\mathrm{Im}\{z_e\}\leq \left(\mathrm{Re}\{z_e\}-\tau_e\right)\tan \Phi,\label{13-constraint1}\\
	&~\mathrm{Im}\{z_e\}\geq \left(\mathrm{Re}\{z_e\}-\tau_e\right)\tan \Phi, \label{13-constraint2}
	\end{align}
\end{subequations}
where  $\mathrm{Re}\{z_e\}$ and $\mathrm{Im}\{z_e\}$ are the real and imaginary part of $z_e$ in the new coordinate system  rotated by the phase of $s_k$.  Since only user 1 has a demand for secure communication, $s_k$ can be specified as $s_1$ when rotating the original coordinate system. Thus, defining $\mathbf{g}_e\triangleq\mathbf{h}_es_1^*$ and introducing the real-valued vectors $\mathbf{a}_e^T=\left[\mathrm{Re}\{\mathbf{g}_e^T\},-\mathrm{Im}\{\mathbf{g}_e^T\}\right]$ and $\mathbf{b}_e^T=\left[\mathrm{Im}\{\mathbf{g}_e^T\},\mathrm{Re}\{\mathbf{g}_e^T\}\right]$, we have
\begin{equation}\label{equation-14}
\mathrm{Re}\{z_e\}=\mathbf{a}_e^T\bar{\mathbf{x}},~\mathrm{Im}\{z_e\}=\mathbf{b}_e^T\bar{\mathbf{x}}.
\end{equation}
Then, the  inequalities in \eqref{equation-13} can be rewritten as 
\begin{subequations} \label{equation-15}
	\begin{align}
	&-\mathbf{b}_e^T\bar{\mathbf{x}}\leq \left(\mathbf{a}_e^T\bar{\mathbf{x}}-\tau_e\right)\tan \Phi,\label{15-constraint1}\\
	&~\mathbf{b}_e^T\bar{\mathbf{x}}\geq \left(\mathbf{a}_e^T\bar{\mathbf{x}}-\tau_e\right)\tan \Phi. \label{15-constraint2}
	\end{align}
\end{subequations}

The goal of the C-D precoding in~\cite{kat1} is to achieve the above  constructive-destructive interference constraints for the legitimate users and the eavesdropper, while minimizing the transmit power. Thus, the optimization problem for the C-D precoding in~\cite{kat1} is given by
\begin{subequations} \label{equation-16}
	\begin{align}
	&\hspace{-4pt}\min_{\bar{\mathbf{x}}}~~~~\|\bar{\mathbf{x}}\|^2\label{eq-16-objective} \\
	&\mbox{s.t.:}~~~~\left(\mathbf{a}_k^T\tan \Phi-\mathbf{b}_k^T\right)\bar{\mathbf{x}}-\tau_0\tan\Phi\geq 0,~\forall k \label{eq-16-constraint-1}\\
	&~~~~~~~~\hspace{1pt}\left(\mathbf{a}_k^T\tan \Phi+\mathbf{b}_k^T\right)\bar{\mathbf{x}}-\tau_0\tan\Phi\geq 0, ~\forall k\label{eq-16-constraint-2}\\
	&~~~~~~~~-\mathbf{b}_e^T\bar{\mathbf{x}}\leq \left(\mathbf{a}_e^T\bar{\mathbf{x}}-\tau_e\right)\tan \Phi, \label{eq-16-constraint-3}\\
	&~~~~~~~~~\mathbf{b}_e^T\bar{\mathbf{x}}\geq \left(\mathbf{a}_e^T\bar{\mathbf{x}}-\tau_e\right)\tan \Phi. \label{eq-16-constraint-4}
	\end{align}
\end{subequations}
Constraints \eqref{eq-16-constraint-1} and \eqref{eq-16-constraint-2} ensure that  the noise-free received signal at each user lies in the  constructive region of the transmitted symbol for that user. Constraints \eqref{eq-16-constraint-3} and \eqref{eq-16-constraint-4} force  the noise-free received signal at the eavesdropper to be located outside the constructive region of $s_1$.  However, it should be noticed that   constraints \eqref{eq-16-constraint-3}  and \eqref{eq-16-constraint-4} correspond to a fraction of the destructive interference region, as shown by the red area in Fig.~\ref{destructive_eve}(b). The {\em  full} destructive region is depicted in red in Fig.~\ref{destructive_eve}(a). The transmit power required to solve the optimization in \eqref{eq-16-objective}-\eqref{eq-16-constraint-4} will thus in general be higher than what would be required if the full destructive region were exploited in formulating the constraints in \eqref{eq-16-constraint-3}  and \eqref{eq-16-constraint-4}.

To achieve a lower transmit power, in the first algorithm we present, we derive a modification to the C-D precoding algorithm that exploits the full destructive region.  As illustrated in Fig.~\ref{destructive_eve}(a),  the entire destructive region can be  described by the  following inequality
\begin{equation}\label{equation-17}
	\left|\mathrm{Im}\{z_e\}\right|\geq \left(\mathrm{Re}\{z_e\}-\tau_e\right)\tan \Phi,
\end{equation}
which is however not convex. The inequality in  \eqref{equation-17} holds when any one of the following three constraints is satisfied
\begin{subequations} \label{equation-18}
	\begin{align}
&\mathrm{Re}\{z_e\}-\tau_e\leq 0, \label{18-constraint1}\\
	&\mathrm{Im}\{z_e\}\geq \left(\mathrm{Re}\{z_e\}-\tau_e\right)\tan \Phi~~\mathrm{and}~~\mathrm{Re}\{z_e\}-\tau_e > 0\label{18-constraint2},\\
	&\!-\mathrm{Im}\{z_e\}\geq \left(\mathrm{Re}\{z_e\}-\tau_e\right)\tan \Phi~~\mathrm{and}~~\mathrm{Re}\{z_e\}-\tau_e > 0\label{18-constraint3}.
	\end{align}
\end{subequations}
The constraints in \eqref{equation-18}  can also be reformulated as a set of simultaneous inequalities using the big-M method~\cite{bigM2}, which requires the introduction of  binary auxiliary variables and thus involves  high-complexity mixed-integer programming. Therefore, for the one-eavesdropper case considered in this paper, we propose to directly use the three constraints in \eqref{equation-18}. The real-valued reformulation of \eqref{equation-18} is given by
\begin{subequations} \label{equation-19}
	\begin{align}
	&~\mathbf{a}_e^T\bar{\mathbf{x}}-\tau_e\leq 0, \label{19-constraint1}\\
	&\left(\mathbf{a}_e^T\tan \Phi-\mathbf{b}_e^T\right)\bar{\mathbf{x}}-\tau_e\tan \Phi\leq 0~~\mathrm{and}~~\mathbf{a}_e^T\bar{\mathbf{x}}-\tau_e > 0\label{19-constraint2},\\
	&\left(\mathbf{a}_e^T\tan \Phi+\mathbf{b}_e^T\right)\bar{\mathbf{x}}-\tau_e\tan \Phi\leq 0~~\mathrm{and}~~\mathbf{a}_e^T\bar{\mathbf{x}}-\tau_e > 0\label{19-constraint3}.
	\end{align}
\end{subequations}

Based on \eqref{equation-19}, our C-D precoding algorithm can be  formulated as
\begin{subequations} \label{equation-20}
	\begin{align}
	& \hspace{-4pt}\min_{\bar{\mathbf{x}}}~~~~\|\bar{\mathbf{x}}\|^2\label{eq-20-objective} \\
	&\mbox{s.t.:}~~~~~\eqref{eq-16-constraint-1}~~~\mathrm{and}~~~\eqref{eq-16-constraint-2} \label{eq-20-constraint-1}\\
	&~~~~~~~~~\eqref{19-constraint1}~~~\mathrm{or}~~~\eqref{19-constraint2}~~~\mathrm{or}~~~\eqref{19-constraint3}  \label{eq-20-constraint-2}.
	\end{align}
\end{subequations}
Note that problem \eqref{equation-20} actually involves three different convex problems, each of which can be readily solved using standard convex optimization solvers. After each subproblem of \eqref{equation-20} is solved, the solution with the minimum transmit power is selected as the final C-D precoder.  In the following  analysis, we will focus on our modified C-D precoder to illustrate the security risk of this type of approach when faced with a smart eavesdropper. A performance comparison between our C-D precoder and the C-D precoder in~\cite{kat1} will be   given in Section~\ref{sec_simulation_results}.

\subsection{Security Risk When Facing  Smart Eavesdropper}\label{sec_compromise_cons_des}

In the above discussions, the eavesdropper is assumed to adopt the same detection method as the legitimate users, namely performing symbol detection  based solely on the instantaneous observed signal. However, if the eavesdropper can exploit  statistical information about the received signal, the symbol detection capability can be  significantly increased.
One  way to obtain statistical information is learning from publicly available training data used for channel estimation.  
On the other hand, when the eavesdropper knows the statistical distribution of all wireless channels, SNR parameters $\gamma_0$ and $\gamma_e$, and the adopted constellation,  she can simulate the  transmission  by solving problem \eqref{equation-20} for different realizations of the wireless channels and   symbols.  
For example, by using either of the above two approaches,   the eavesdropper can  empirically determine the  probability distribution of the received signal conditioned on each symbol $c_m\in \mathcal{C}_M$. Note that for  PSK modulation, it is sufficient to focus only on the phase of the received signal, which is denoted as $\theta_e$.  We denote the  conditional PDF of $\theta_e$ given symbol $c_m$ as $f(\theta_e|c_m)$. With empirically derived estimates of these PDFs, the eavesdropper can employ the ML criterion for symbol detection. For the ML detector, when the eavesdropper receives a new signal with phase $\theta_{e,0}$, the detection result is  given by
\begin{equation}\label{equation-21}
c_m^* = \mathop {\arg \max }\limits_{c_m} ~f(\theta_{e,0}|c_m).
\end{equation}
Note that the ML detector in \eqref{equation-21} not only utilizes the instantaneous information $\theta_{e,0}$ but also the statistical information about the distribution of the phase. We refer to an eavesdropper using this ML detector as  a smart eavesdropper.

\begin{figure}[t]
	\centering
	\subfloat[]{
		\includegraphics[width=2.5in]{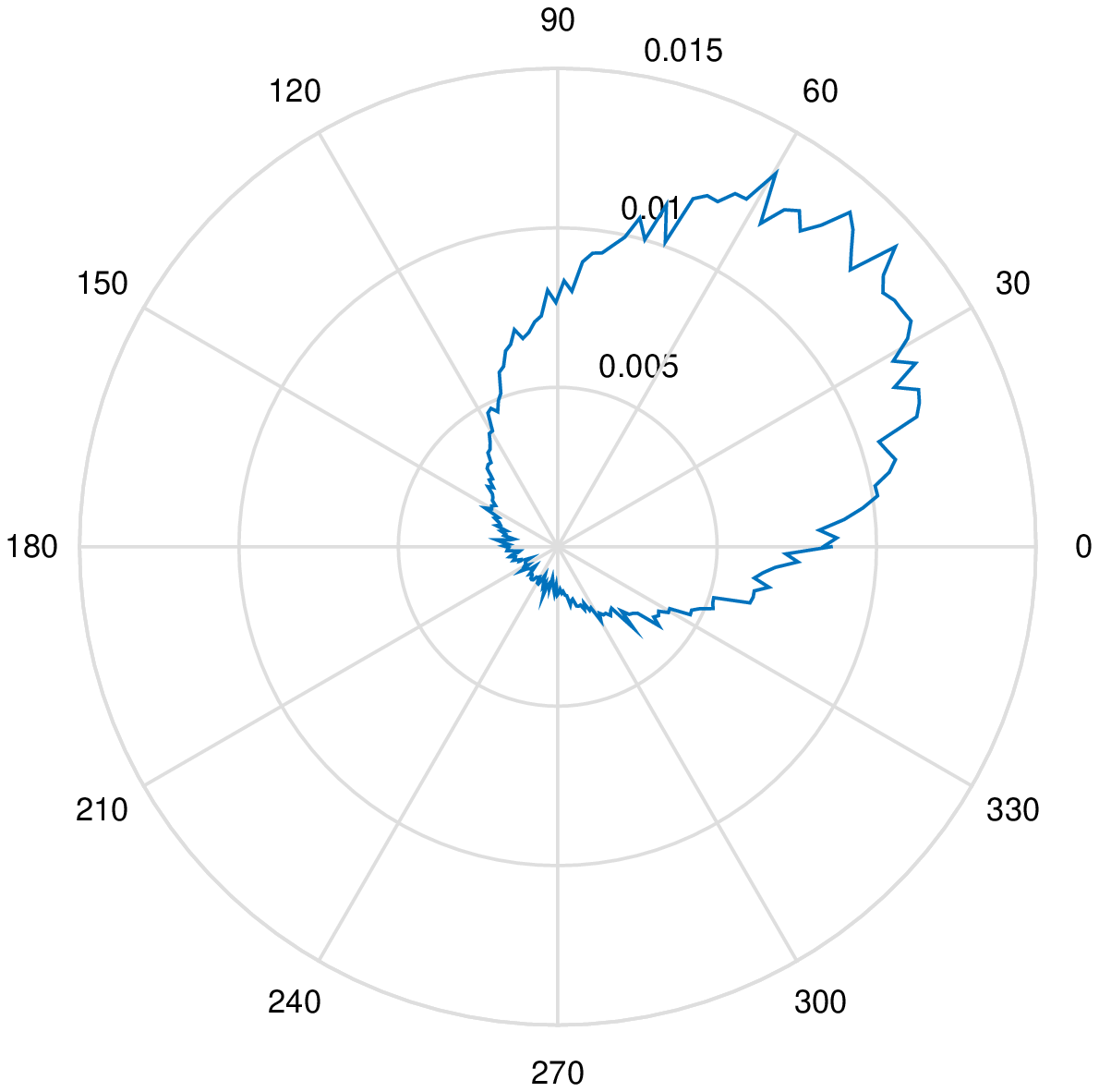}
	}
	\hfil
	\subfloat[]{
		\includegraphics[width=2.5in]{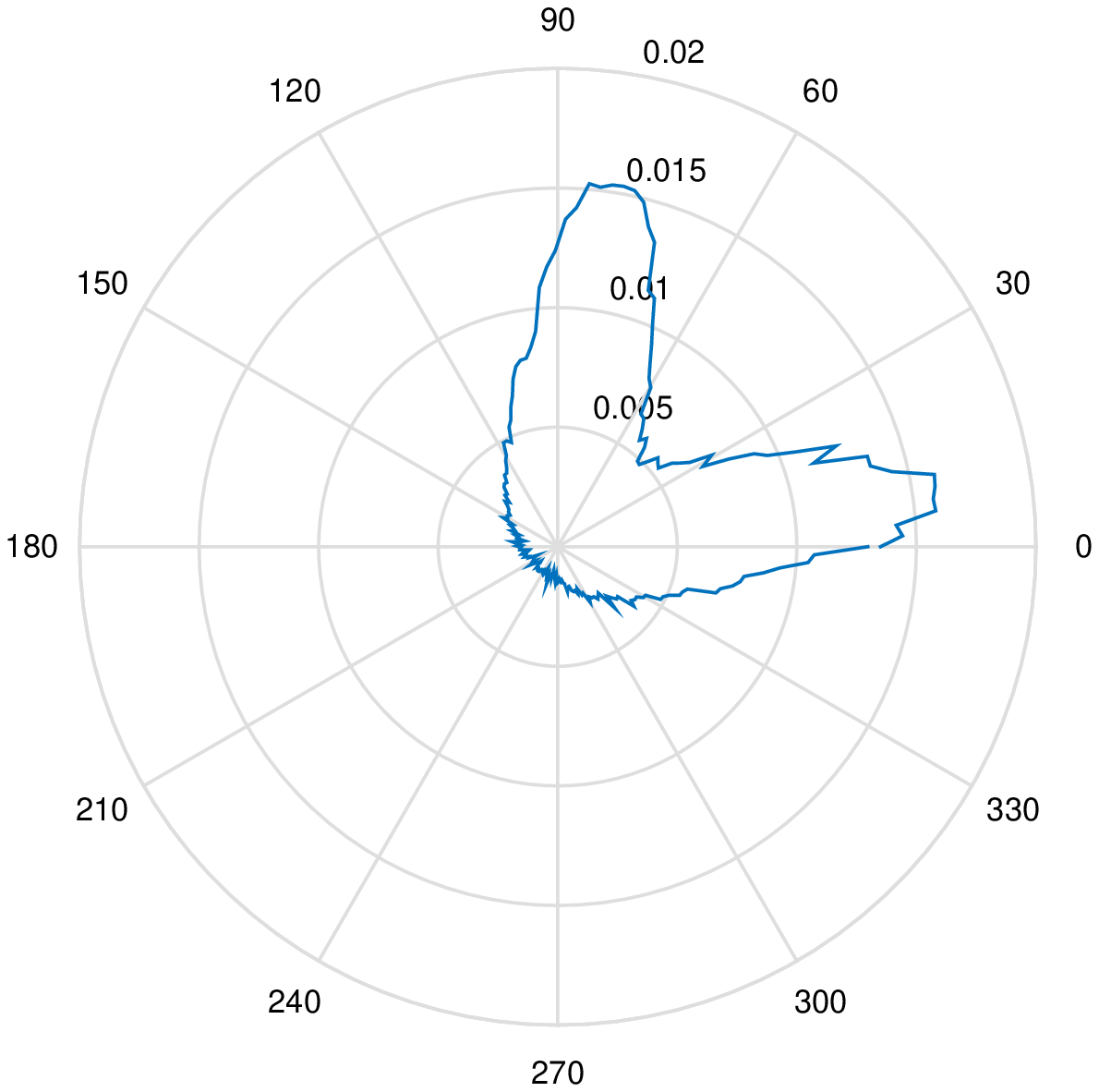}
	}
	\hfil 
	\subfloat[]{
		\includegraphics[width=2.5in]{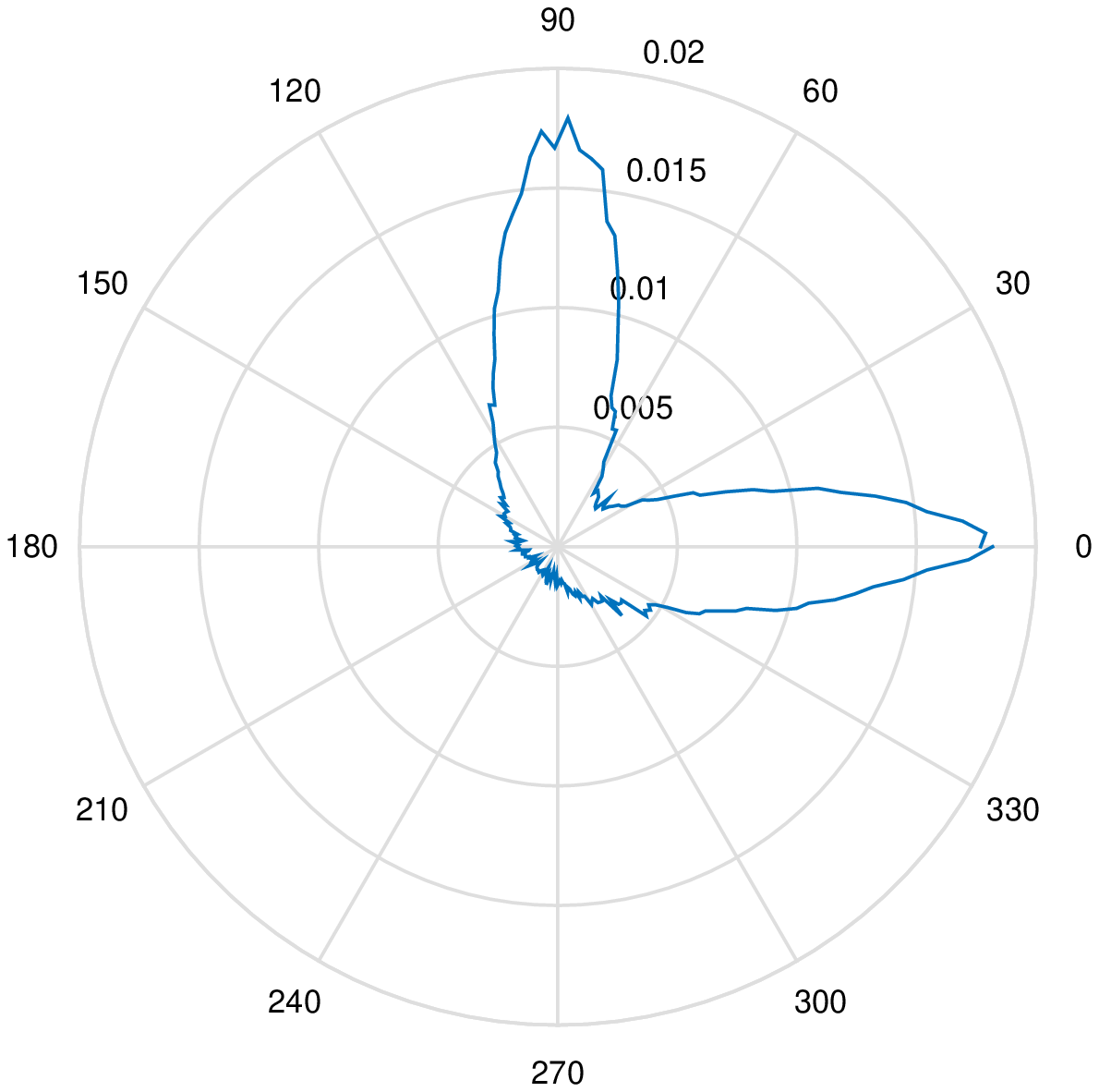}
	}
	\caption{ QPSK modulation with channel correlation coefficient $\rho=0.3$ and $\gamma_0=10~\mathrm{dB}$: Empirical conditional PDF $f(\theta_e|c_1)$ for different SNR requirements at the eavesdropper.  (a) $\gamma_e=15~\mathrm{dB}$  (b) $\gamma_e=0~\mathrm{dB}$  (c) $\gamma_e=-15~\mathrm{dB}$.
	}
	\label{kat_phase_QPSK}
\end{figure}

To study the performance of C-D precoding in the presence of a smart eavesdropper, we examine empirically estimated distributions $f(\theta_e | c_1)$ for a few scenarios\footnote{Considering that symbol $c_m$ is  a rotated version of $c_1$ by angle  $(m-1)\frac{\ang{360}}{M}$, the empirical conditional PDF $f(\theta_e|c_m)$ which has been averaged over  the transmitted symbols for other users and all wireless channels  is also a rotated version of $f(\theta_e|c_1)$ by  the same angle. Therefore, the empirical conditional PDFs for other symbols have characteristics  similar  to $f(\theta_e|c_1)$. }.  In Fig.~\ref{kat_phase_QPSK} we show $f(\theta_e|c_1)$  for  QPSK  modulation in a polar coordinate system  where the polar angle from  \ang{0} to \ang{360} covers  the  range of values for $\theta_e$ and the radius represents  probability density. The results are obtained by averaging over both the transmit symbol vector $\mathbf{s}$ where user 1's desired symbol is fixed as $c_1$,   and all wireless channels with $\beta_k=1~(k=1,\cdots,K)$ and $\beta_e=1$.  

First, as illustrated in Fig.~\ref{kat_phase_QPSK}(a), when  $\gamma_e$ is very large which leads to nearly  no constraint on the  eavesdropper's received signal, there is a considerable probability that the eavesdropper's received signal  lies in the decision region of $c_1$ (i.e., quadrant I). As such, the eavesdropper has a relatively high probability of detecting the transmitted symbol, and thus it is important to adopt security-aware precoding when channel correlation exists. 
As $\gamma_e$ becomes smaller, we can see that the C-D algorithm produces a distribution for $f(\theta_e|c_1)$ with two narrow lobes\footnote{As shown later in Fig.~\ref{realkat_phase_QPSK}, if the original approach of~\cite{kat1} were used, the lobe would  be along only one of the symbol boundaries rather than symmetrically on both of them.}. Compared with Fig.~\ref{kat_phase_QPSK}(b), the lobes in Fig.~\ref{kat_phase_QPSK}(c)  are pushed closer to the  boundaries of the destructive region of $c_1$,  i.e., the two dotted black rays in Fig.~\ref{destructive_eve}(a). This is because compared with pushing $z_e$ deeper into the destructive region, pushing $z_e$  near the boundaries can save power in most cases, which is consistent with the   objective  of problem \eqref{equation-20}.   As a result,  the probability that the  eavesdropper's received signal falls in quadrant I is reduced, as illustrated in   Fig.~\ref{kat_phase_QPSK}(b) and Fig.~\ref{kat_phase_QPSK}(c). This is beneficial for communication secrecy if the eavesdropper adopts the same simple detector as the legitimate user. 

However, the rotational asymmetry of $f(\theta_e | c_1)$ when $\gamma_e$ is small can be exploited by the eavesdropper to improve her detection probability. The distributions for different $c_m$ will be identical to those shown in Fig. 4, except rotated by $\pm 90^\circ$ and $180^\circ$ for the case of QPSK. The lobes for adjacent symbols will overlap, and the correct detection probability for the eavesdropper can approach 50\% for equally probable symbols, twice the minimum of 25\% achieved by random guessing. Thus, whether $\gamma_e$ is large or small, the security of the C-D precoding scheme is compromised. This means that the security of the system may not improve even as the transmit power increases or even if the desired user's channel is much stronger than that of the eavesdropper. Clearly, a better approach for a smart eavesdropper employing optimal ML detection would be to ensure that the distribution for different $f(\theta_e | c_m)$ are as identical and rotationally invariant as possible. This is the theme of the technique proposed in the next section.

\section{Secure Precoding  against Smart Eavesdropper} \label{sec_secure_precoding_smart_eavesdropper}
 
In this section, we first discuss a general principle for  designing secure precoding against smart eavesdroppers. Then, we present a specific precoding algorithm based on information-carrying signal suppressing  (ICSS),  which can  overcome the drawback of the  C-D precoding scheme. A fast  ICSS precoding algorithm is also provided, which  uses the gradient projection  method for  improving computational  efficiency.
 
\subsection{Design Principle} \label{sec_design_principle}

We know that for \emph{M}-PSK modulation, the eavesdropper's minimum correct detection probability is $1/M$, which is the  probability achieved by  a random guess.  To make random guessing the optimal strategy for a smart eavesdropper, the  conditional PDFs for different symbols should be identical.  Furthermore, since the conditional PDFs for other symbols are rotated versions of $f(\theta_e|c_1)$, the   ideal $f(\theta_e|c_1)$ which makes the transmitted symbol totally indistinguishable from other symbols  should be rotationally symmetric for every angle of  $\ang{360}/M$, as mentioned in Section~\ref{sec_compromise_cons_des}. 
Therefore, to mitigate smart eavesdropping, the focus of  secure precoding is not on  the instantaneous performance but on making the received symbols statistically indistinguishable.
 
\subsection{ICSS Precoding} \label{sec_SNR_Constrained}

Based on the above discussion,  secure precoding against a smart eavesdropper should aim at making the  conditional PDFs for different symbols identical, which can be specified as making the conditional PDF rotationally symmetric in a polar coordinate system.  The simplest type of rotational symmetry would be to make the conditional PDFs uniform in all directions, i,e, where the phase $\theta_e$ is uniformly distributed on $[0,2\pi]$. However, even for this special case it is still challenging to design a specific precoding algorithm that can realize the desired statistical distribution.

\begin{figure}[t]
	\centering
	\subfloat[]{
		\includegraphics[width=2.3in]{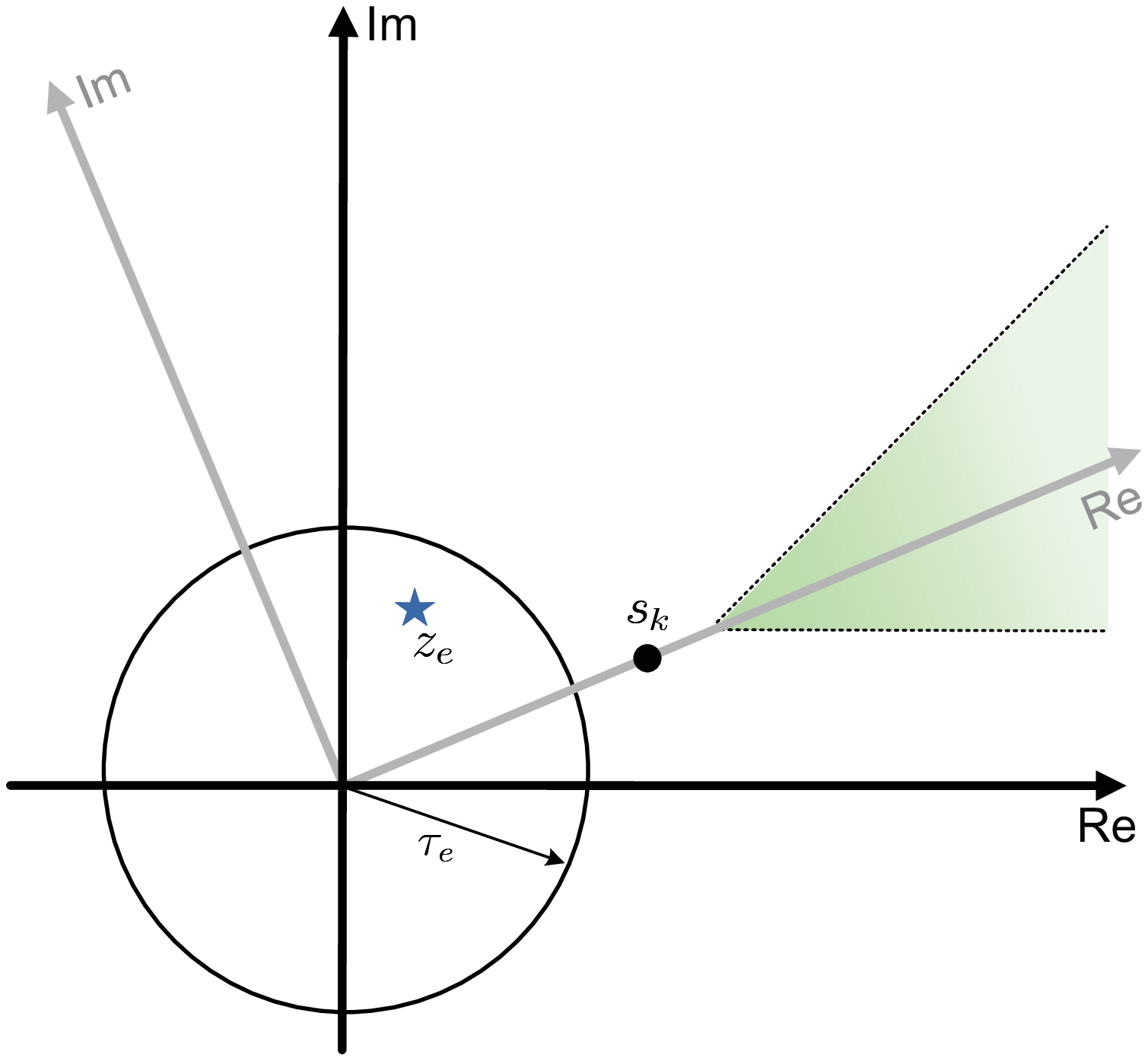}
	}
	\hfil 
	\subfloat[]{
		\includegraphics[width=2.3in]{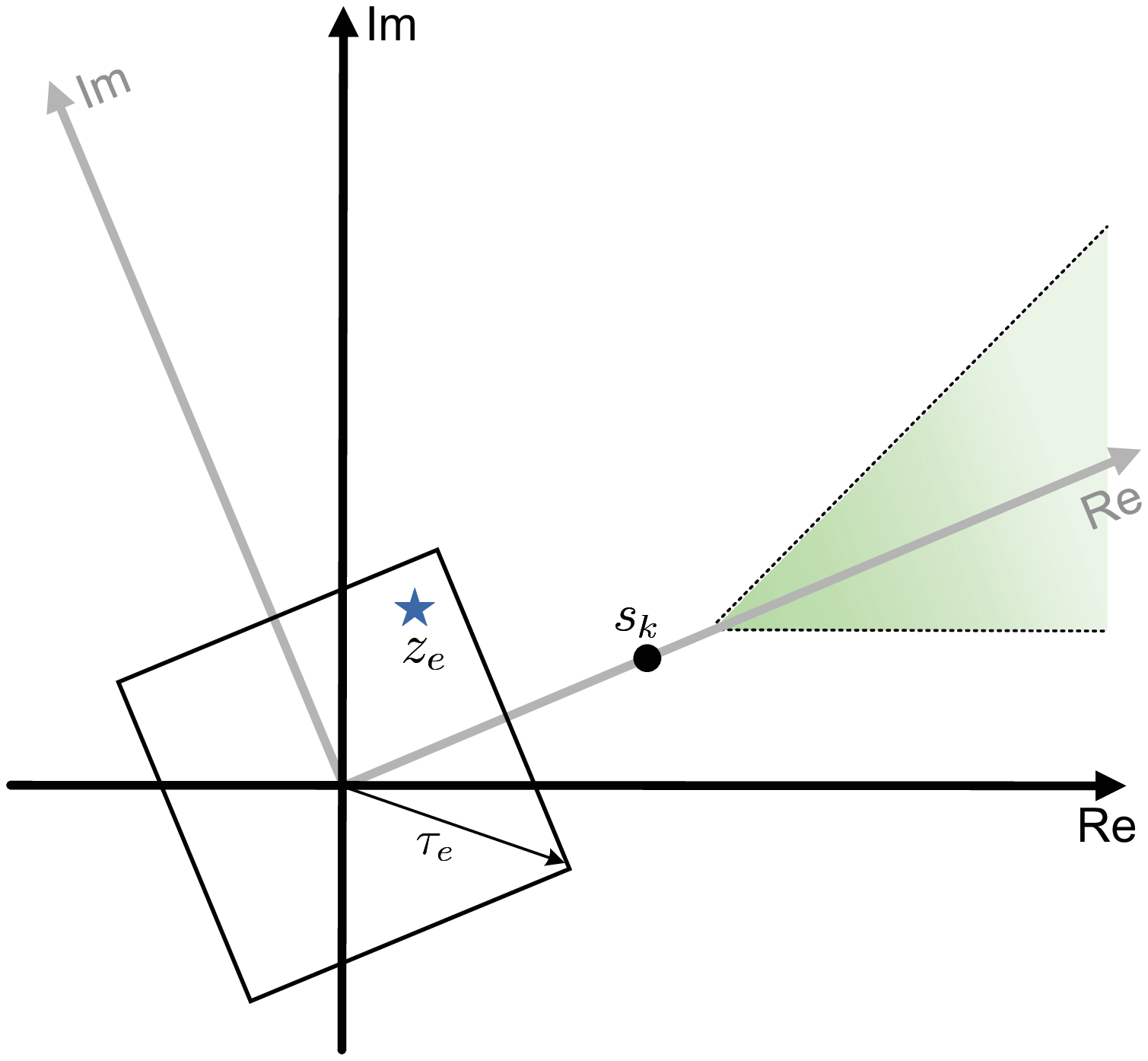}
	}
	\caption{  Secure precoding against smart eavesdropper, where $\tau_e=\sqrt{\gamma_e}$: (a) ICSS  precoding (b) Fast ICSS precoding.
	}
	\label{SP_precoding}
\end{figure}

It is worth noting that  eavesdropper's received signal consists of two parts, the information-carrying signal $z_e$ which is also the aforementioned  noise-free received signal, and the additive Gaussian noise $n_e$ whose phase is uniformly distributed on $[0,2\pi]$. If the information-carrying signal can be hidden in the additive noise,  the ML detector at the eavesdropper will be useless. This can be achieved by simply transmitting in the null space of the eavesdropper channel. However, this will consume a lot of power especially when the eavesdropper channel is highly correlated with the main channel. To show the trade-off between consumed power and secrecy, we consider a more general and flexible approach   which limits the power of the information-carrying signal, as illustrated in Fig.~\ref{SP_precoding}(a).  Following the previous notation, for  a given SNR constraint $\gamma_e$, the  power of $z_e$ is limited as
\begin{equation}\label{equation-22}
|z_e|^2 \leq \tau_e^2,
\end{equation}
where $\tau_e=\sqrt{\gamma_e}$. For the complex  variable $z_e$ we have $|z_e|^2 = \left(\mathrm{Re}\{z_e\}\right)^2+\left(\mathrm{Im}\{z_e\}\right)^2$. Therefore, the real-valued reformulation of \eqref{equation-22} is given by
\begin{equation}\label{equation-23}
\bar{\mathbf{x}}^T\left(\mathbf{a}_e\mathbf{a}_e^T+\mathbf{b}_e\mathbf{b}_e^T\right)\bar{\mathbf{x}}\leq \tau_e^2.
\end{equation}

Based on \eqref{equation-23}, our proposed ICSS precoding approach can be stated as
\begin{subequations} \label{equation-24}
	\begin{align}
	&\hspace{-4pt}\min_{\bar{\mathbf{x}}}~~~~\|\bar{\mathbf{x}}\|^2\label{eq-24-objective} \\
	&\mbox{s.t.:}~~~~\left(\mathbf{a}_k^T\tan \Phi-\mathbf{b}_k^T\right)\bar{\mathbf{x}}-\tau_0\tan\Phi\geq 0,~\forall k \label{eq-24-constraint-1-1}\\
	&~~~~~~~~\left(\mathbf{a}_k^T\tan \Phi+\mathbf{b}_k^T\right)\bar{\mathbf{x}}-\tau_0\tan\Phi\geq 0, ~\forall k \label{eq-24-constraint-1-2}\\
	&~~~~~~~~~\bar{\mathbf{x}}^T\mathbf{A}\bar{\mathbf{x}}\leq \tau_e^2 \label{eq-24-constraint-2},
	\end{align}
\end{subequations}
where $\mathbf{A}=\mathbf{a}_e\mathbf{a}_e^T+\mathbf{b}_e\mathbf{b}_e^T$. Constraints \eqref{eq-24-constraint-1-1} and \eqref{eq-24-constraint-1-2} ensure that the constructive region requirement  is met at each desired user while constraint \eqref{eq-24-constraint-2} limits the maximum power of the eavesdropper's noise-free received signal. 
Problem \eqref{equation-24} is convex  and can be solved using standard algorithms.  

Note that although  $z_e$ is referred to as the information-carrying signal, it can be any point on the complex plane, rather than one of the fixed  constellation points.  By placing a constraint on the power of $z_e$, we actually limit the contribution that $z_e$ makes to the final received signal. As  the power of $z_e$ decreases, the information about the real transmitted symbol  is more likely to be hidden. Therefore, the achievable secrecy level of our ICSS precoding approach will continue improving  as $\gamma_e$ decreases, which overcomes the drawback of  C-D precoding where secrecy performance can not be further improved as $\gamma_e$ is reduced.

\subsection{Fast ICSS  Precoding} \label{sec_fast_SNR_Constrained}

In the  last subsection, we proposed an ICSS precoding algorithm that improves communication security by suppressing the power of the information-carrying signal at the eavesdropper.  However, the power constraint in \eqref{eq-24-constraint-2} leads to a quadratically constrained quadratic program~\cite{S.Boyd}, the computational complexity of which is relatively high. To further reduce the computational complexity, we change the power constraint  in \eqref{eq-24-constraint-2} to the power constraint  illustrated in Fig.~\ref{SP_precoding}(b).  By requiring $z_e$ to be located in a square with diagonal length $2\tau_e$, the maximum power of $z_e$ is $\tau_e^2$. The advantage of this approach is that the maximum power constraint can be expressed using linear rather than quadratic constraints:
\begin{subequations} \label{equation-25}
	\begin{align}
	&-\sqrt{2}\tau_e/2\leq\mathrm{Re}\{z_e\}\leq \sqrt{2}\tau_e/2, \label{25-constraint1}\\
	&-\sqrt{2}\tau_e/2\leq\mathrm{Im}\{z_e\}\leq \sqrt{2}\tau_e/2\label{25-constraint2}.
	\end{align}
\end{subequations}
Based on the real-valued expression \eqref{equation-14}, the fast ICSS precoding algorithm can thus be formulated as
\begin{subequations} \label{equation-26}
	\begin{align}
	&\hspace{-4pt}\min_{\bar{\mathbf{x}}}~~~~\|\bar{\mathbf{x}}\|^2\label{eq-26-objective} \\
	&\mbox{s.t.:}~~~~~\mathbf{Q}\bar{\mathbf{x}}-\mathbf{b} \preceq \mathbf{0}, \label{eq-26-constraint-1}
	\end{align}
\end{subequations}
where
\begin{align} \label{equation-27}
&\mathbf{Q}=\left[{\begin{array}{*{20}{c}}
	{-\mathbf{a}_k^T\tan \Phi+\mathbf{b}_k^T}\\
	{-\mathbf{a}_k^T\tan \Phi-\mathbf{b}_k^T}\\	
	{\mathbf{a}_e^T}\\
	{-\mathbf{a}_e^T}\\
	{\mathbf{b}_e^T}\\
	{-\mathbf{b}_e^T}\\
	\end{array}}\right]_{(2K+4)\times 2N},\nonumber\\
&~\mathbf{b}=\left[{\begin{array}{*{20}{c}}
	{-\tau_0\tan\Phi\mathbf{1}_{2K\times 1}}\\
	{\frac{\sqrt{2}}{2}\tau_e\mathbf{1}_{4\times 1}}
	\end{array}}\right]_{(2K+4)\times 1}.
\end{align}

To solve problem \eqref{equation-26}, we first derive the following Lagrange dual function of  \eqref{equation-26}
\begin{equation} \label{equation-28}
L(\bar{\mathbf{x}},\bm{\lambda})=\bar{\mathbf{x}}^T\bar{\mathbf{x}}+\bm{\lambda}^T\left(\mathbf{Q}\bar{\mathbf{x}}-\mathbf{b}\right),
\end{equation}
where $\bm{\lambda}\succeq \mathbf{0}$ is the $(2K+4)\times 1$ Lagrange multiplier  associated with  constraint \eqref{eq-26-constraint-1}. According to  the Karush-Kuhn-Tucker (KKT) conditions~\cite{S.Boyd}, the optimal solution to problem \eqref{equation-26} is  achieved at $\frac{\partial L}{\partial \bar{\mathbf{x}}}=\mathbf{0}$. The root of  $\frac{\partial L}{\partial \bar{\mathbf{x}}}=\mathbf{0}$ is given by
\begin{equation} \label{equation-29}
\bar{\mathbf{x}}^*=-\frac{1}{2}\mathbf{Q}^T\bm{\lambda}.
\end{equation}
Then, we need to find the optimal $\bm{\lambda}$, which  is the solution to the dual problem given by
\begin{equation} \label{equation-30}
\max_{\bm{\lambda}\succeq \mathbf{0}}~ ~g(\bm{\lambda}),
\end{equation}
 where 
\begin{equation} \label{equation-31}
g(\bm{\lambda})= \inf~ L(\bar{\mathbf{x}},\bm{\lambda})\mathop{=}\limits^{(a)}-\frac{1}{4}\bm{\lambda}^T\mathbf{Q}\mathbf{Q}^T\bm{\lambda}-\bm{\lambda}^T\mathbf{b}
\end{equation}
is the dual function, and  (a) is obtained by plugging \eqref{equation-29} into \eqref{equation-28}. 

Due to the non-negative constraint $\bm{\lambda}\succeq \mathbf{0}$, it is difficult to derive a closed-form solution to \eqref{equation-30}. Therefore, we use  the gradient projection  method to find the solution.   Since we aim at maximizing the concave function $g(\bm{\lambda})$, the current value $\bm{\lambda}_n$ is updated as
\begin{equation} \label{equation-32}
\bm{\lambda}_{n+1}=\max\left\{\bm{\lambda}_n+t_n\!\bigtriangledown \!g(\bm{\lambda}_n),\mathbf{0}\right\},
\end{equation}
where   $t_n$ is the positive step size used at the $n$th iteration, and $\bigtriangledown g(\bm{\lambda})=-\frac{1}{2}\mathbf{Q}\mathbf{Q}^T\bm{\lambda}-\mathbf{b}$ is the gradient of $g(\bm{\lambda})$. The function $\max \left\{\bm{\lambda},\mathbf{0}\right\}$ is the projection of $\bm{\lambda}$ onto the  solution space $\bm{\lambda}\succeq \mathbf{0}$ in \eqref{equation-30}. 
For calculating the step size $t_n$ in each iteration, we employ the backtracking line search algorithm~\cite{S.Boyd,Jiang}. The  details are shown in  Algorithm~1, where the  parameters $t$, $\delta$, and $\mu$  are set according to~\cite{Jiang}.  
With the  updated step size, we can use the gradient projection method to solve problem \eqref{equation-30} and then problem \eqref{equation-26}. The entire iterative algorithm is summarized in Algorithm~2.

\begin{algorithm}\label{alg_backtracking}
	\setstretch{1}
	\caption{Backtracking line search algorithm.} 
	{\bf Input:} $\mathbf{Q}$, $\mathbf{b}$, $\bm{\lambda}_n$, $t =1$, $\delta=0.1$, $\mu=0.5$;\!\!\!\!\!\!\!\!\!\!\!\!\!\!\!\!\!\!\!\!\!\!\!\!
	\begin{algorithmic}[1]
		%\algsetup{linenosize=\footnotesize}
		%\small
		\WHILE{1} 
		\STATE  $\bm{\lambda}_{n+1}=\max \left\{\bm{\lambda}_n+t*\bigtriangledown g(\bm{\lambda}_n),\mathbf{0}\right\}$;
		\IF{$g(\bm{\lambda}_{n+1})\geq g(\bm{\lambda}_n)+\delta\cdot \bigtriangledown g(\bm{\lambda}_n)^T(\bm{\lambda}_{n+1}-\bm{\lambda}_n)$}
		\STATE Break;
		\ENDIF
		\STATE  $t=\mu t$;
		\ENDWHILE
		\STATE $t_n = t;$
	\end{algorithmic}
	 {\bf Output:} $t_n$
\end{algorithm}

\begin{algorithm}\label{alg_gradient}
	\setstretch{1}
	\caption{Iterative algorithm for solving problem \eqref{equation-26}.} 
	{\bf Input:} 
	$\mathbf{Q}$, $\mathbf{b}$, initial $\bm{\lambda}_0\succeq \mathbf{0}$, $n=-1$, $\epsilon$;\!\!\!\!\!\!\!\!\!\!\!\!\!\!\!\!\!\!\!\!\!\!\!\!
	\begin{algorithmic}[1]
		\REPEAT
		\STATE $n=n+1$;
		\STATE Calculate step size $t_n$ using Algorithm~1;
		\STATE $\bm{\lambda}_{n+1}=\max\left\{\bm{\lambda}_n+t_n\!\bigtriangledown \!g(\bm{\lambda}_n),\mathbf{0}\right\}$;	
		\UNTIL $g(\bm{\lambda}_{n+1})-g(\bm{\lambda}_n)\leq \epsilon$ or maximum number of iterations reached
		\STATE Obtain the optimal Lagrange multiplier, i.e., $\bm{\lambda}^*=\bm{\lambda}_{n+1}$;
	\end{algorithmic}
   {\bf Output:} $\bar{\mathbf{x}}^*=-\frac{1}{2}\mathbf{Q}^T\bm{\lambda}^*$
\end{algorithm}

Now, we compare the computational complexity of the ICSS precoding and the fast ICSS precoding algorithms. For  ICSS precoding, problem \eqref{equation-24} can be reformulated as a semidefinite programming problem with  linear matrix inequality (LMI) constraints, which can be solved using the interior-point method (IPM). According to~\cite{complexity}, the computational complexity in each iteration of IPM is $\mathcal{O}\left(N^4\right)$. On the other hand, as discussed in~\cite{Jiang}, the dominant complexity of Algorithm~2  is calculating the gradient $\bigtriangledown g(\bm{\lambda})$ in each iteration, which requires $\mathcal{O}\left(KN\right)$ computations. Therefore, the fast ICSS precoding algorithm is significantly less complicated.

\section{Secure Precoding without Eavesdropper's CSI} \label{sec_noCSI}

In this section, we consider the scenario where the eavesdropper's CSI is unavailable\footnote{The scenario without eavesdropper's CSI is also  studied in~\cite{kat1} but only for the case of a single legitimate user. Moreover, the effect of  channel correlation is not considered in~\cite{kat1}. }. Without the eavesdropper's CSI, we cannot control the location of $z_e$  and thus we cannot control the distribution of the eavesdropper's  received signal. On the other hand, due to the channel correlation, it is very likely that the eavesdropper's received signal is aligned with or near the symbol of interest, which makes the symbol easily  detected. %Fortunately, it is still possible to achieve secure transmission as long as the eavesdropper's and the legitimate user's channels are not identical. 
Recalling the channel correlation model in \eqref{equation-4}, the noise-free received signal at the eavesdropper can be decomposed as
\begin{equation} \label{equation-33}
z_e=\mathbf{h}_e^T\mathbf{x}=\underbrace{\sqrt{\beta_e/\beta_1}\sqrt{\rho}\hspace{1pt} \mathbf{h}_1^T\mathbf{x}}_{\mathrm{symbol~of~interest}}+\underbrace{\sqrt{\beta_e(1-\rho)}\hspace{1pt}\mathbf{w}^T\mathbf{x}}_{\mathrm{random~term}}.
\end{equation}
The first term in \eqref{equation-33} represents the desired symbol for user 1, since $\mathbf{h}_1^T\mathbf{x}$ is designed to lie in the constructive region of $s_1$. The second term in \eqref{equation-33} is unknown and random, changing  every transmission due to different $\mathbf{x}$.  Since the value of $\mathbf{w}$ is unknown, the random term $\mathbf{w}^T\mathbf{x}$ can be either constructive or destructive to the instantaneous detection of  $s_1$. This  term randomizes the statistical distribution of $\theta_e$ and thus can degrade the detection performance of the  eavesdropper. Therefore, the second term in  \eqref{equation-33} acts as a type of  AN as in~\cite{Goel,Zhou} which helps hide the transmitted information.

To be  hidden  in the AN term,  the power of $\mathbf{h}_1^T\mathbf{x}$ in  \eqref{equation-33} should be as small as possible. However,  the constraint  in \eqref{equation-7}  requires that $z_1=\mathbf{h}_1^T\mathbf{x}$ be located in the constructive region of $s_1$. It is obvious that the power-minimizing point in a  constructive region is the intersection  of the two constructive region boundaries. Consequently, we have $	\mathrm{Re}\{z_1\}=\tau_0$ and $\mathrm{Im}\{z_1\}=0$, which is equivalent to
\begin{equation} \label{equation-34}
\mathbf{a}_1^T\bar{\mathbf{x}}=\tau_0,~\mathbf{b}_1^T\bar{\mathbf{x}}=0.
\end{equation}
In addition, the power of $\mathbf{w}^T\mathbf{x}$ should be large enough to guarantee sufficient randomization to the phase of the eavesdropper's received signal. With   unknown $\mathbf{w}$, we place a constraint on the power of $\mathbf{x}$ instead, since the average power of $\mathbf{w}^T\mathbf{x}$ can be approximated as $\mathbb{E}_\mathbf{w}\left\{ |\mathbf{w}^T\mathbf{x}|^2\right\}\approx \Tr\left( \mathbf{x}^H \mathbb{E}_\mathbf{w}\left\{ \mathbf{w}^*\mathbf{w}^T\right\}\mathbf{x}\right)=\|\mathbf{x}\|^2=\|\bar{\mathbf{x}}\|^2$.

Based on the above discussion, we propose the following approach to be used when CSI for the potential eavesdropper is absent:
\begin{subequations} \label{equation-35}
	\begin{align}
	&\hspace{-4pt}\min_{\bar{\mathbf{x}}}~~~~\|\bar{\mathbf{x}}\|^2\label{eq-35-objective} \\
	&\mbox{s.t.:}~~~~~\mathbf{a}_1^T\bar{\mathbf{x}}=\tau_0,~\mathbf{b}_1^T\bar{\mathbf{x}}=0, \label{eq-35-constraint-1}\\
	&~~~~~~~~\left(\mathbf{a}_k^T\tan \Phi-\mathbf{b}_k^T\right)\bar{\mathbf{x}}-\tau_0\tan\Phi\geq 0,~\forall k\neq 1 \label{eq-35-constraint-2}\\
	&~~~~~~~~\hspace{1pt}\left(\mathbf{a}_k^T\tan \Phi+\mathbf{b}_k^T\right)\bar{\mathbf{x}}-\tau_0\tan\Phi\geq 0 , ~\forall k\neq 1 \label{eq-35-constraint-3}\\
	&~~~~~~~~~\|\bar{\mathbf{x}}\|^2\geq P_0, \label{eq-35-constraint-4}
	\end{align}
\end{subequations}
where $P_0$ determines the minimum average power of the  AN. Without constraint \eqref{eq-35-constraint-4}, problem \eqref{equation-35} only focuses on  minimizing $\|\bar{\mathbf{x}}\|^2$, which  may lead to  small AN values. Constraint \eqref{eq-35-constraint-2} and \eqref{eq-35-constraint-3} are the  constructive region requirements  for the other users. 

Constraint \eqref{eq-35-constraint-4} leads to a non-convex feasible set, which makes problem \eqref{equation-35} difficult to solve. To tackle this problem, we adopt the sequential convex programming (SCP) method~\cite{SCP1,SCP2}, which iteratively approximates the original non-convex set by an inner convex one in each iteration.  Specifically, in the $n$th iteration we approximate function $f(\bar{\mathbf{x}})\triangleq\|\bar{\mathbf{x}}\|^2$  using its first-order Taylor expansion $\tilde{f}(\bar{\mathbf{x}},\bar{\mathbf{x}}_n)$ at the current point $\bar{\mathbf{x}}_n$. The expression for $\tilde{f}(\bar{\mathbf{x}},\bar{\mathbf{x}}_n)$ is given by
\begin{align} \label{equation-36}
\tilde{f}(\bar{\mathbf{x}},\bar{\mathbf{x}}_n)&=f(\bar{\mathbf{x}}_n)+\bigtriangledown f(\bar{\mathbf{x}}_n)^T\left(\bar{\mathbf{x}}-\bar{\mathbf{x}}_n\right)\nonumber \\
&=\|\bar{\mathbf{x}}_n\|^2+2\bar{\mathbf{x}}_n^T\left(\bar{\mathbf{x}}-\bar{\mathbf{x}}_n\right),
\end{align}
where $\bigtriangledown f(\bar{\mathbf{x}})$ is the gradient of $f(\bar{\mathbf{x}})$ with respect to $\bar{\mathbf{x}}$. The constraint \eqref{eq-35-constraint-4} can now be replaced with
\begin{equation} \label{equation-37}
\tilde{f}(\bar{\mathbf{x}},\bar{\mathbf{x}}_n) \geq P_0.
\end{equation}
With \eqref{equation-37}, we can obtain the following convex optimization problem in the  $n$th iteration 
\begin{subequations} \label{equation-38}
	\begin{align}
	&\hspace{-4pt}\min_{\bar{\mathbf{x}}}~~~~\|\bar{\mathbf{x}}\|^2\label{eq-38-objective} \\
	&\mbox{s.t.:}~~~~~\!\eqref{eq-35-constraint-1}-\eqref{eq-35-constraint-3},~\eqref{equation-37} \label{eq-38-constraint-1}.
	\end{align}
\end{subequations}
Problem \eqref{equation-38} can be readily solved using standard convex optimization solvers or the gradient projection method in Section~\ref{sec_fast_SNR_Constrained}.

\begin{algorithm}\label{alg_SCP}
	\caption{SCP method for solving problem \eqref{equation-35}.} 
	{\bf Input:} 
	initial point $\bar{\mathbf{x}}_0$, $n=0$, $\epsilon$;
	\begin{algorithmic}[1]
		\REPEAT
		\STATE Calculate  function \eqref{equation-36} based on $\bar{\mathbf{x}}_n$;
		\STATE Solve problem \eqref{equation-38} and assign the optimal solution to $\bar{\mathbf{x}}_{n+1}$;
		\STATE $n=n+1$;
		\UNTIL $ \|\bar{\mathbf{x}}_{n-1}\|^2-\|\bar{\mathbf{x}}_n\|^2 \leq \epsilon$ or maximum number of iterations reached
	\end{algorithmic}
	 {\bf Output:} $\bar{\mathbf{x}}_n$
\end{algorithm}

As  shown in~\cite{SCP2}, the SCP method results in an objective function that is non-increasing at every iteration, so this approach will always converge to a local optimum of problem (36).  We summarize the entire SCP method  in Algorithm~3.
To guarantee that the generated solution in each iteration is feasible for the original non-convex problem  \eqref{equation-35}, we need to start from a point $\bar{\mathbf{x}}_0$ which  belongs to the feasible set of problem \eqref{equation-35}.  To find $\bar{\mathbf{x}}_0$, we use the iterative feasibility search  algorithm (IFSA) proposed in~\cite{SCP2}, which iteratively minimizes the  violation parameter until convergence.

\section{Simulation Results} \label{sec_simulation_results}

In this section, we compare the performance of the above precoding algorithms through simulation. Unless otherwise stated, we set $N=6$, $K=3$,  $\beta_k=1~(k=1,\cdots,K)$, $\beta_e=1$,  and $\gamma_0=10~\mathrm{dB}$. The eavesdropper in the simulation  below adopts the  detection method in \eqref{equation-21}. To simulate the performance of the eavesdropper, we first generate  random transmit symbols and wireless channels to obtain the empirical conditional PDF $f(\theta_e|c_m)$. Then,  the empirical decision region for each constellation point $c_m$ can be obtained, which is the region where $f(\theta_e|c_m)$ is larger than  $f(\theta_e|c_n)$ for any $n \neq m$.  Based on the decision region,  the  detection probability for the eavesdropper can then be  obtained.  The transmit power in the results below is also averaged over different transmit symbols and wireless channels. In the following simulations, we first consider the scenario where  the eavesdropper's CSI is available. Then, the scenario without eavesdropper's CSI is studied.

\begin{figure}[t]
	\centering
	\subfloat[]{
		\includegraphics[width=2.5in]{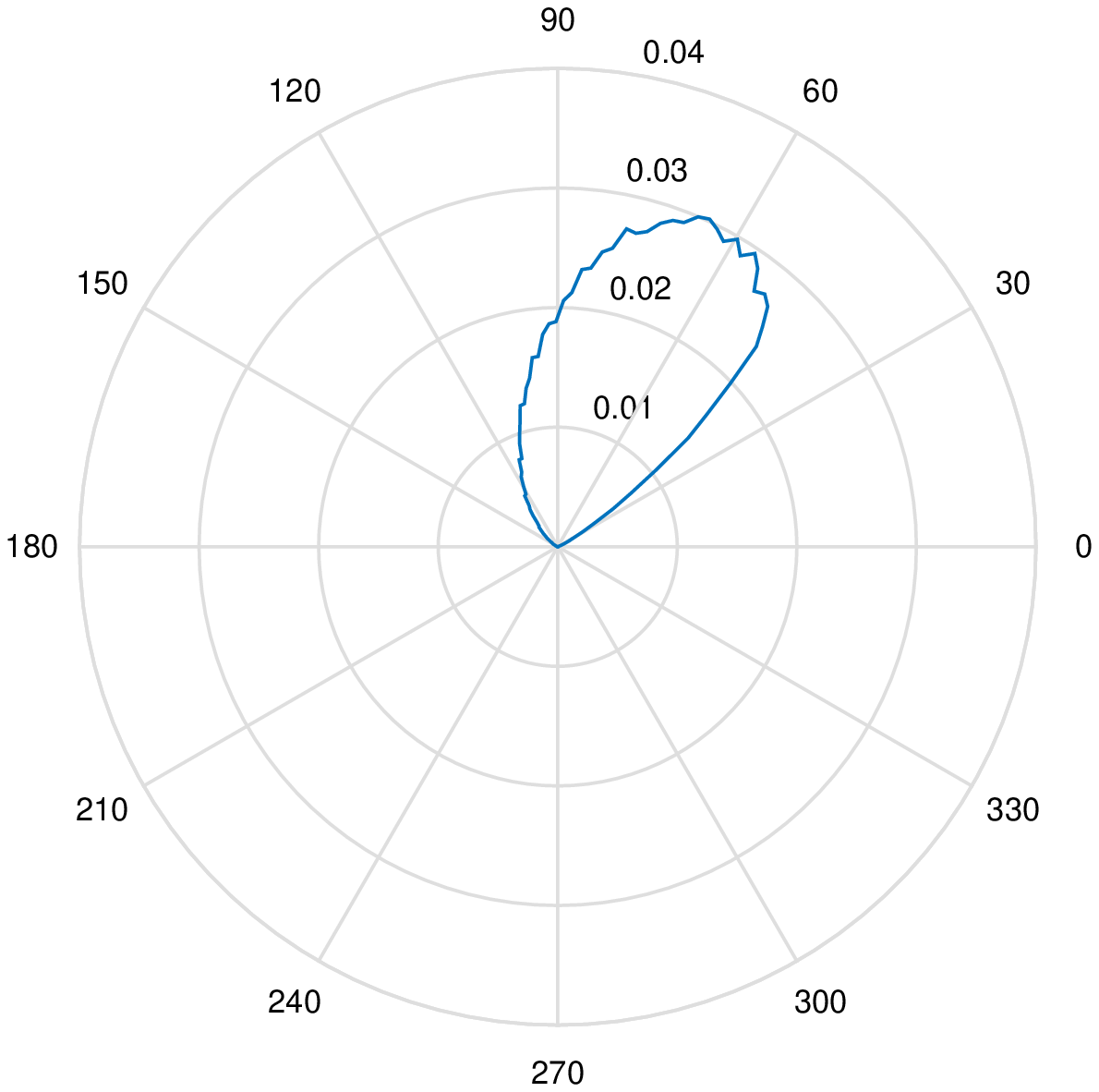}
	}
	\hfil
	\subfloat[]{
		\includegraphics[width=2.5in]{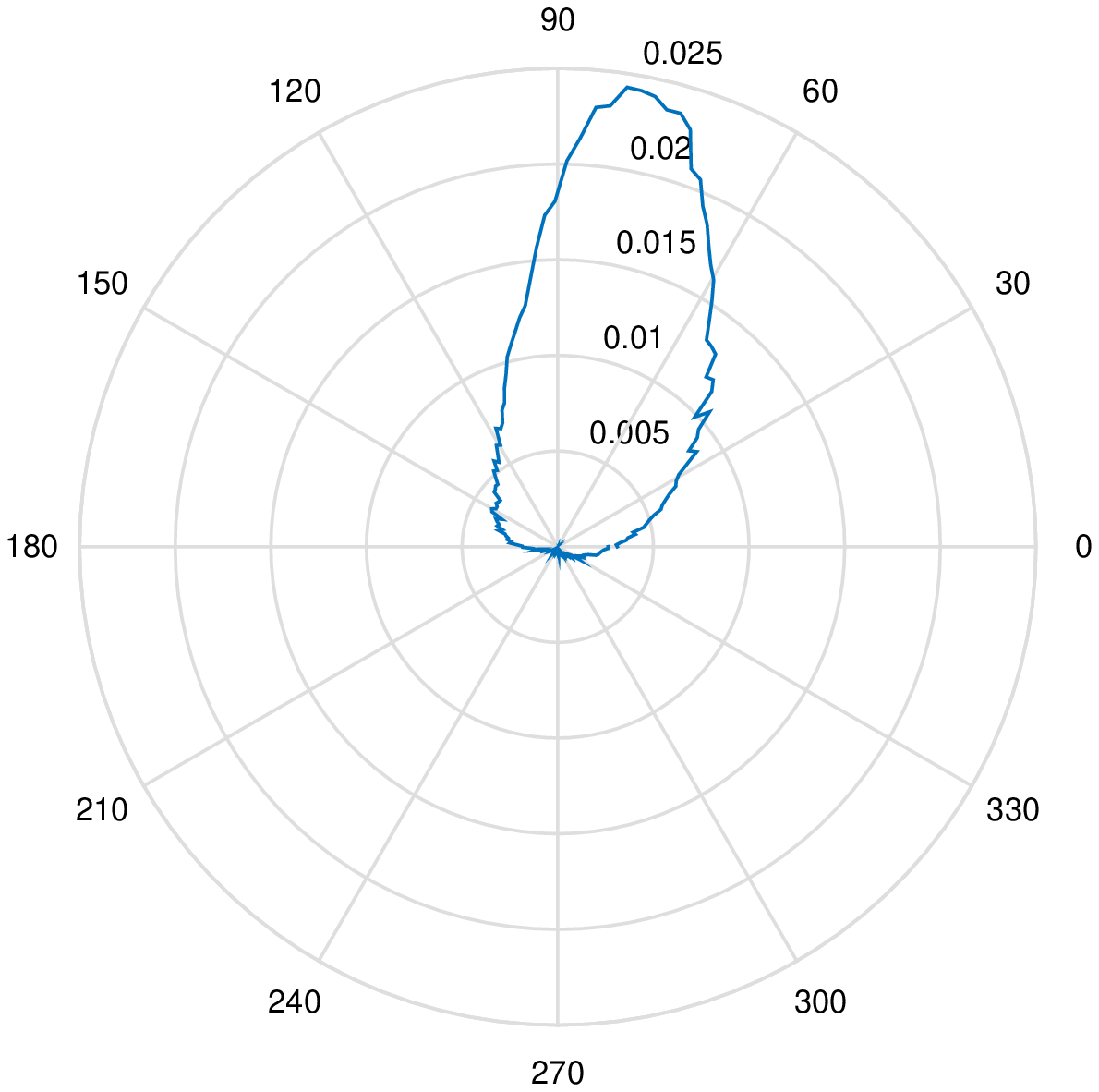}
	}
	\hfil 
	\subfloat[]{
		\includegraphics[width=2.5in]{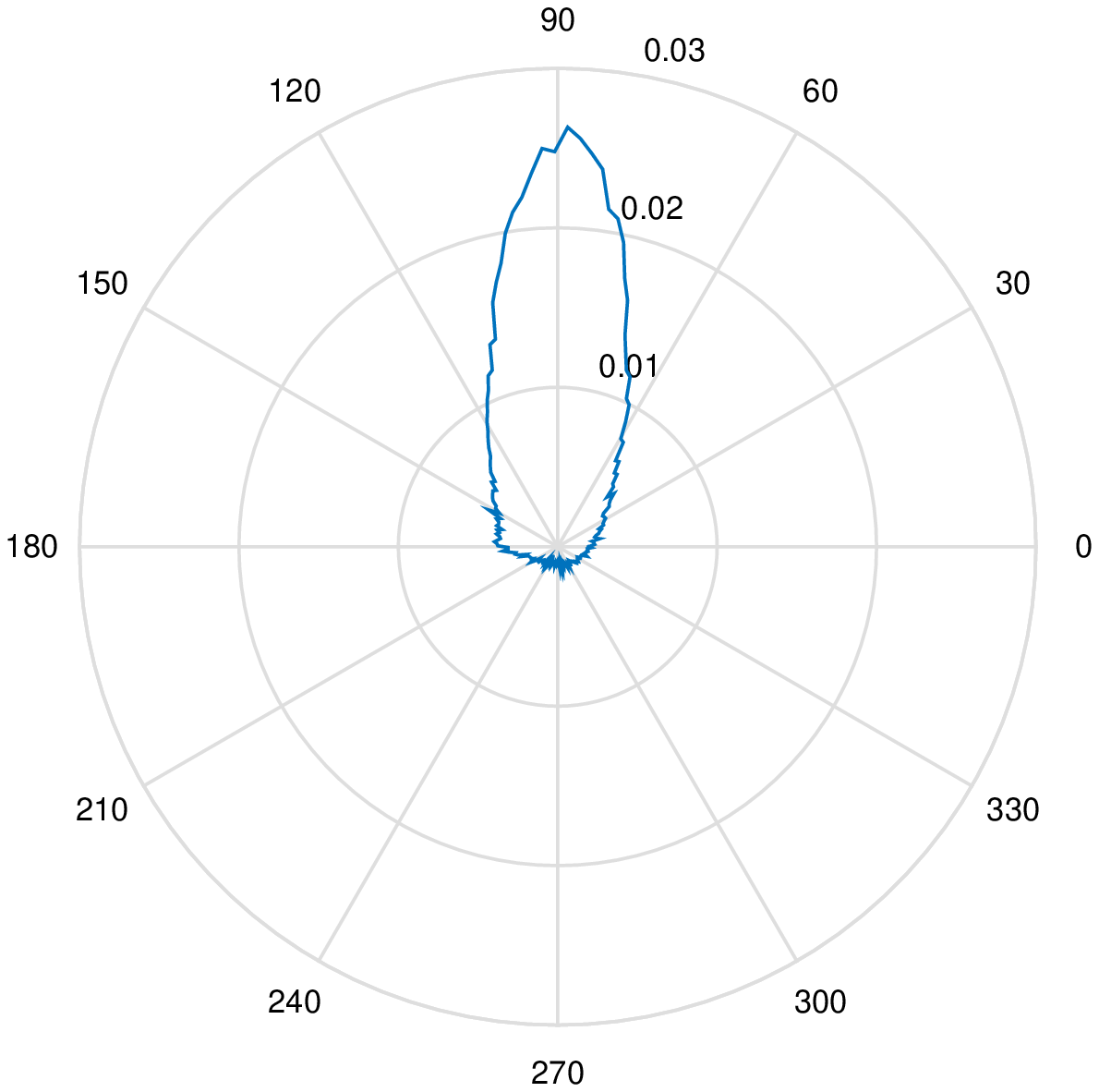}
	}
	\caption{ The C-D precoding algorithm in~\cite{kat1} with QPSK modulation: Empirical conditional PDF $f(\theta_e|c_1)$ for different SNR requirements at the eavesdropper.  (a) $\gamma_e=15~\mathrm{dB}$  (b) $\gamma_e=0~\mathrm{dB}$  (c) $\gamma_e=-15~\mathrm{dB}$.
	}
	\label{realkat_phase_QPSK}
\end{figure}

Similar to Fig.~\ref{kat_phase_QPSK}, we first depict the empirical conditional PDF $f(\theta_e|c_1)$   for  the C-D precoding algorithm in~\cite{kat1}, where QPSK modulation is adopted with channel correlation parameter $\rho=0.3$, a desired SNR of $\gamma_0=10~\mathrm{dB}$.  Unlike Fig.~\ref{kat_phase_QPSK}, we can see that there is only one lobe in Fig.~\ref{realkat_phase_QPSK}, since the C-D precoding in~\cite{kat1}  requires $\mathrm{Im}\{z_e\}\geq 0$, which only exploits the upper part of the entire destructive region. Moreover, when $\gamma_e$ is larger than $\gamma_0$, e.g., $\gamma_e=15~\mathrm{dB}$ in Fig.~\ref{realkat_phase_QPSK}(a), there is still a significant lobe which makes the transmitted symbol easily detected. This is because the feasible region shown by the red zone in Fig.~\ref{destructive_eve}(b)  shrinks as $\gamma_e$ increases. Thus, the phase distribution of $z_e$ is still concentrated rather than dispersed.

\begin{figure}[t]
	\centering
	\subfloat[]{
		\includegraphics[width=3.3in]{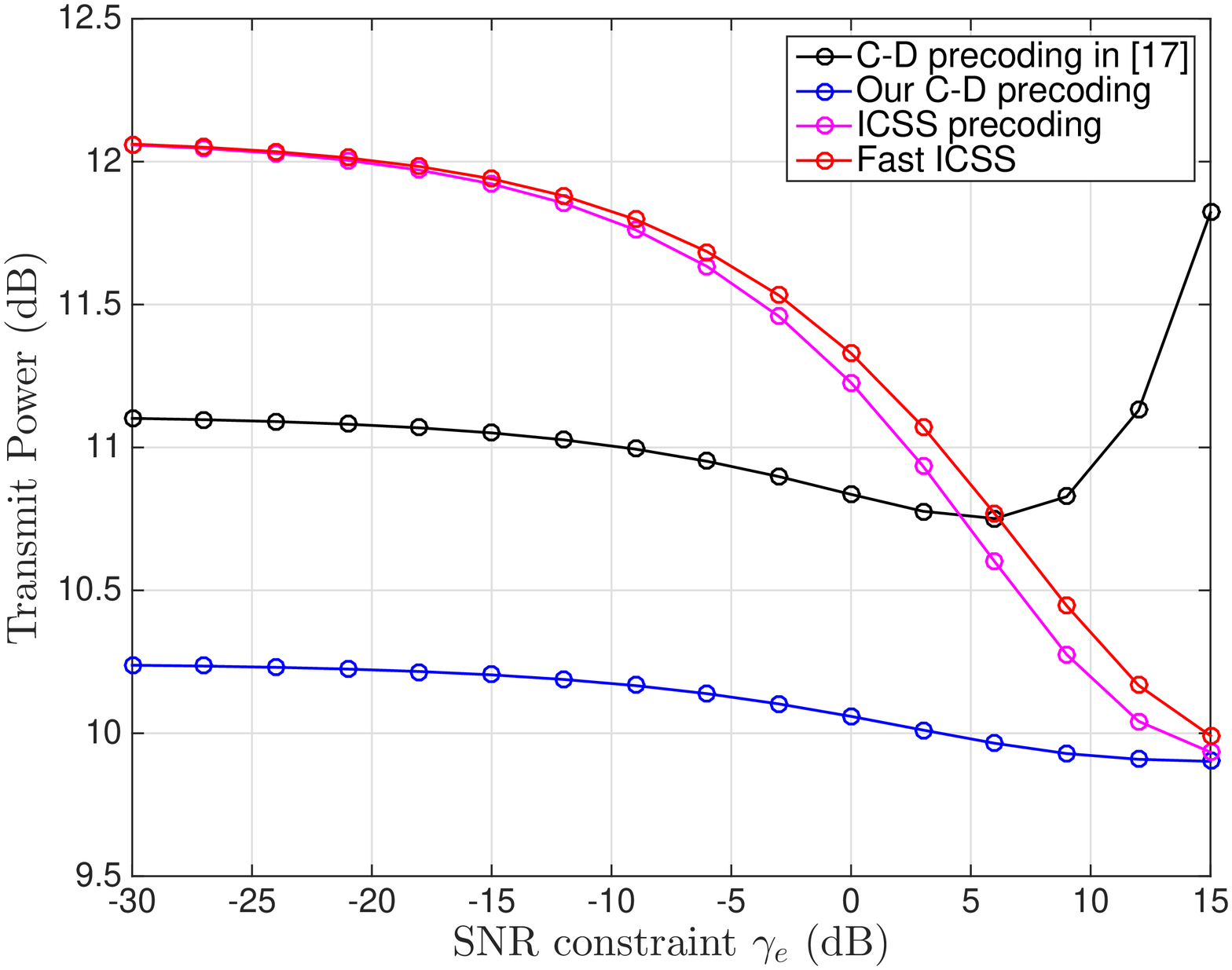}
	}
	\hfil
	\subfloat[]{
		\includegraphics[width=3.3in]{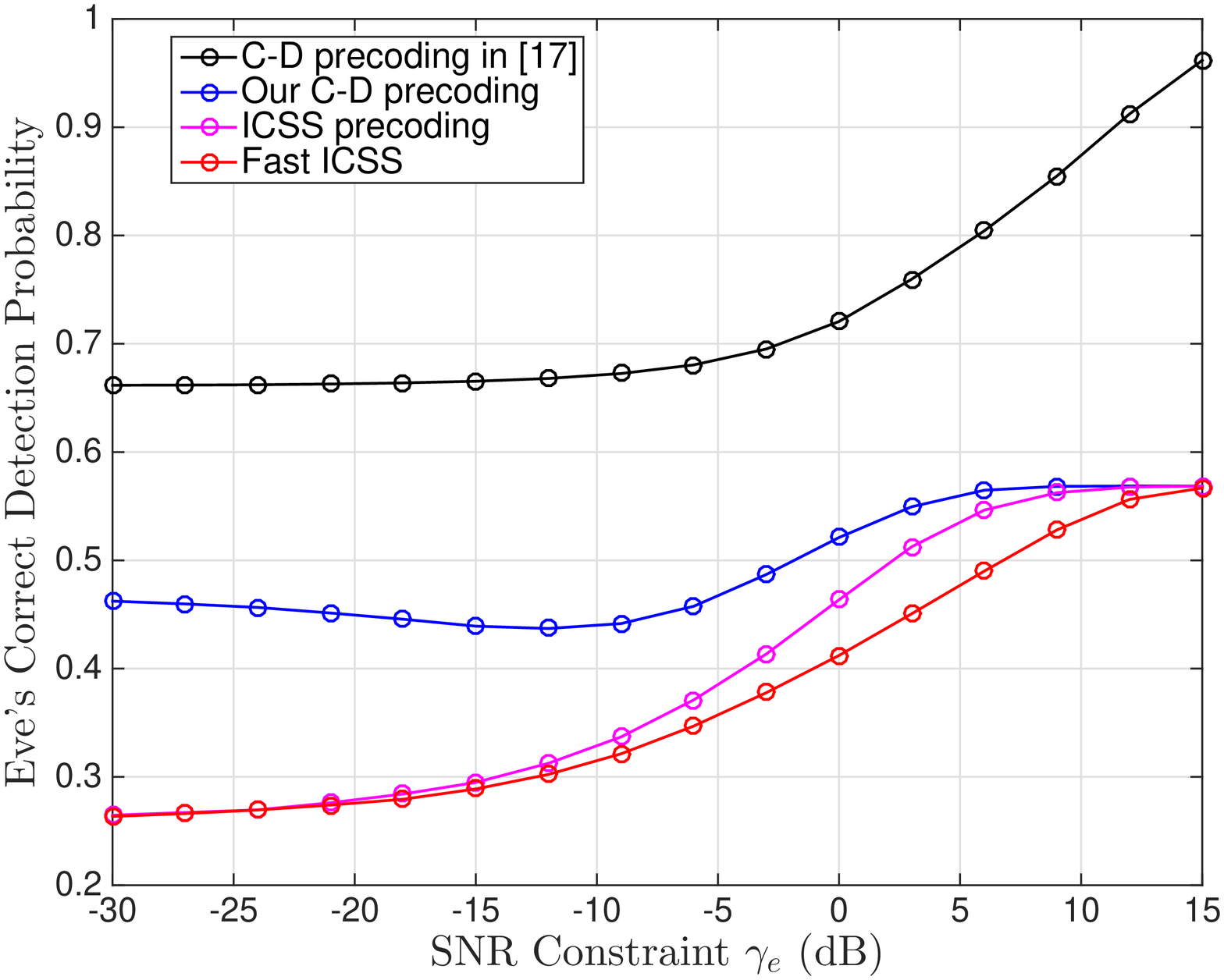}
	}
	\caption{ Performance comparison among different precoding algorithms for QPSK modulation with $\rho=0.3$.  (a) Transmit power versus SNR constraint  (b) Eavesdropper's correct detection probability versus SNR constraint.
	}
	\label{powerandprob_compare}
\end{figure}

In Fig.~\ref{powerandprob_compare}, we compare the algorithms discussed in the paper  in terms of consumed power and secrecy. First, we concentrate on the comparison between the two C-D precoding strategies. From Fig.~\ref{powerandprob_compare}(a) we can see that the C-D precoding  in~\cite{kat1} consumes more power than our C-D precoding algorithm. This is expected since the scheme in~\cite{kat1}  only exploits a part of the destructive region, which leads to a stricter optimization.  In addition, it is worth noting that the  power consumed by our C-D precoding algorithm decreases as $\gamma_e$ increases,  while the power of the scheme in~\cite{kat1} first decreases and then increases.  This is because the feasible set for our approach keeps getting larger as $\gamma_e$ increases. However, for the algorithm in~\cite{kat1},  too large a $\gamma_e$ will unnecessarily
require the imaginary part of $z_e$ to be  large, which consumes significant power\footnote{Typically,  $\gamma_e$ is chosen to be smaller than  $\gamma_0$ to degrade the eavesdropper's performance. Here we consider a wider range of $\gamma_e$ to thoroughly study the effect of $\gamma_e$ on the performance of each precoding algorithm.}. For the secrecy performance in Fig.~\ref{powerandprob_compare}(b), our C-D precoding also outperforms the C-D precoding in~\cite{kat1}.  This corresponds to the result in Fig.~\ref{realkat_phase_QPSK} that a single lobe makes the transmitted symbol more easily  detected.

\begin{figure}
	\centering
	\includegraphics[width=3.3in]{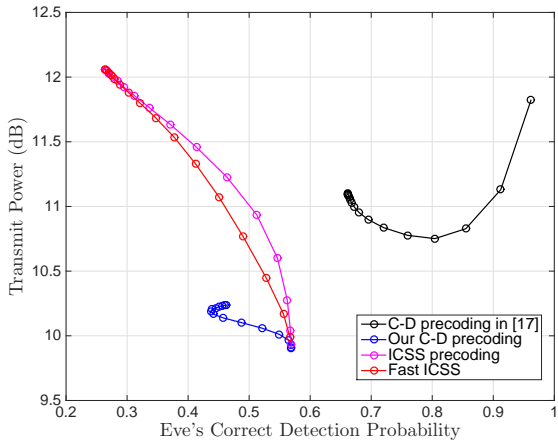}
	\caption{ Power-secrecy trade-off curve for QPSK modulation  with channel correlation coefficient $\rho=0.3$.}
	\label{Tradeoff_QPSK3}
\end{figure}

Next, we compare the performance of the two C-D precoding algorithms and the two ICSS algorithms.   Fig.~\ref{powerandprob_compare}(b) shows that our improved C-D precoding strategy provides significantly lower detection probability than the algorithm of~\cite{kat1}, which achieves a probability of slightly less than 0.5 for small $\gamma_e$.  On the other hand, the ICSS and fast ICSS  schemes can further reduce the eavesdropper's detection probability to about 0.25,  which corresponds to the perfect secrecy case where the eavesdropper's optimal strategy is random guessing. However, as shown by  Fig.~\ref{powerandprob_compare}(a), the consumed power of the two ICSS algorithms for $\gamma_e<5~\mathrm{dB}$ is higher than that of the two C-D precoding schemes, particularly our  C-D precoding algorithm. Therefore, there is a trade-off between secrecy and power, which will be discussed later in Fig.~\ref{Tradeoff_QPSK3}. 

\renewcommand{\arraystretch}{1.25} 
\begin{table*}[tp]  	
	\centering  
	\fontsize{8}{8}\selectfont  
	\begin{threeparttable}  
		\caption{Performance comparison between  ICSS precoding ($\gamma_e=-30~\mathrm{dB}$) and the traditional ZF precoding.}  
		\label{tab:performance_comparison}  
		\begin{tabular}{ccccccc}  
			\toprule  
			\multirow{3}{*}{Algorithm}&  
			\multicolumn{3}{c}{ $N=6$,~$K=3$}&\multicolumn{3}{c}{ $N=4$,~$K=3$}\cr  
			\cmidrule(lr){2-4} \cmidrule(lr){5-7}  
			&Power (dB)&$P_\mathrm{dec}^\mathrm{Eve}$&$\overline{\mathrm{SER}}_k$& Power (dB)&$P_\mathrm{dec}^\mathrm{Eve}$&$\overline{\mathrm{SER}}_k$\cr  
			\midrule  
			ICSS&12.06&0.26&$1.4\times 10^{-3}$&22.34&0.26&$1.4\times 10^{-3}$\cr  
			ZF&12.34&0.25&$1.5\times 10^{-3}$&27.50&0.25&$1.5\times 10^{-3}$\cr  
			\bottomrule  
		\end{tabular}  \label{ZF}
	\end{threeparttable}  
\end{table*}  

\renewcommand{\arraystretch}{0.65}

Now we focus on the extreme case for the ICSS precoding algorithm where  $\gamma_e=-30~\mathrm{dB}$.  In this case, the information-carrying signal $z_e$ at the eavesdropper is nearly zero, which corresponds to the traditional zero-forcing (ZF) scheme where both the MUI and the eavesdropper's received signal are forced to be zero. Table~\ref{ZF} makes a comparison between the ICSS precoding algorithm ($\gamma_e=-30~\mathrm{dB}$) and the traditional ZF algorithm for QPSK modulation with  $\rho=0.3$, where $P_\mathrm{dec}^\mathrm{Eve}$ denotes the eavesdropper's   detection probability, and $\overline{\mathrm{SER}}_k$ is the symbol error rate averaged over  $K$ legitimate users. From Table~\ref{ZF}  one can see that   the two algorithms achieve  similar detection performance at the legitimate users and the eavesdropper, but  ICSS precoding requires less power especially when the system is heavily loaded (e.g., a 5dB power savings for the case of $N=4$,~$K=3$).  Unlike the ZF method whose consumed power is fixed and high, the ICSS algorithm can adjust its transmit power according to the secrecy requirement. Even for the scenario with a high secrecy requirement, ICSS precoding can use much less power to achieve  secrecy performance similar to the ZF method, especially for a heavily loaded  system.

\begin{figure}[t]
	\centering
	\subfloat[]{
		\includegraphics[width=3.3in]{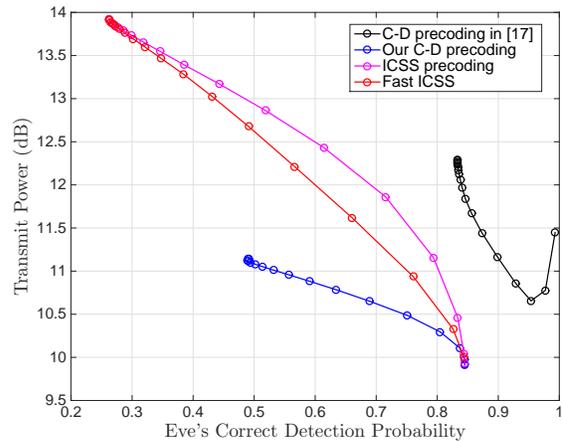}
	}
	\hfil
	\subfloat[]{
		\includegraphics[width=3.3in]{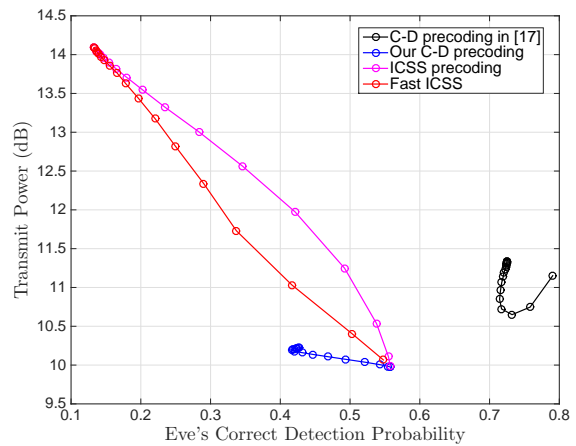}
	}
	\caption{ Power-secrecy trade-off curve for a highly correlated scenario with $\rho=0.7$.  (a) QPSK modelation  (b) 8PSK modulation.
	}
	\label{highcorrelation_compare}
\end{figure}

Fig.~\ref{Tradeoff_QPSK3} depicts the power-secrecy trade-off curves for the two C-D precoding algorithms and the two ICSS algorithms, which are generated by changing the SNR constraint $\gamma_e$ at the eavesdropper. First, it can be observed that the C-D precoding in~\cite{kat1} has the worst performance, since all the other schemes  use less power but achieve a lower detection probability at the eavesdropper. In addition, our C-D precoding algorithm  is the most energy efficient  when  the eavesdropper's detection probability is higher than 0.4. However, it is worth noting that this scheme as well as the scheme in~\cite{kat1} cannot further degrade eavesdropper's performance, which indicates that there is a secrecy bottleneck for the C-D precoding algorithms. Further increases in the transmit power for our C-D precoding algorithm enhance the eavesdropper's detection probability. Therefore, the two ICSS precoding schemes are vital for eliminating the secrecy bottleneck of the C-D precoding algorithm. It is also worth noting that for the simulation scenario in  Fig.~\ref{Tradeoff_QPSK3}, the fast ICSS precoding is more energy efficient than  ICSS precoding. Recalling Fig.~\ref{powerandprob_compare} which has the same simulation settings, it is observed that  for a fixed $\gamma_e$, although  the fast ICSS scheme consumes more power than the ICSS algorithm, it can significantly reduce the eavesdropper's detection probability.   Consequently,  fast ICSS precoding has a better trade-off curve than  ICSS precoding. One possible reason is that compared with using a circle to limit the location of $z_e$, using a square which has a side perpendicular to the line on which the  desired symbol  is located  can more effectively reduce the probability that the eavesdropper's received signal is aligned with or near the desired symbol.

In Fig.~\ref{highcorrelation_compare}, we study the  trade-off between power and secrecy  in an environment with higher channel correlation. Comparing Fig.~\ref{highcorrelation_compare}(a) and Fig.~\ref{Tradeoff_QPSK3}, we can see that to achieve the same secrecy level, more power is needed for the  scenario with $\rho=0.7$. Moreover,  when the correlation parameter rises from 0.3 to 0.7, the lowest  detection probability for the eavesdropper that the two C-D precoding schemes can achieve is increased. On the contrary, the lowest probability that the two ICSS precoding algorithms can achieve remains unchanged, as long as there is enough power. Therefore,  ICSS precoding is more robust to the  strength of the channel correlation.  We also study the case of 8PSK modulation in Fig.~\ref{highcorrelation_compare}(b). It is observed that the two ICSS precoding algorithms can reduce the  eavesdropper's detection probability to about 0.125, which corresponds to random guessing for 8PSK.

\begin{figure}
	\centering
	\includegraphics[width=3.3in]{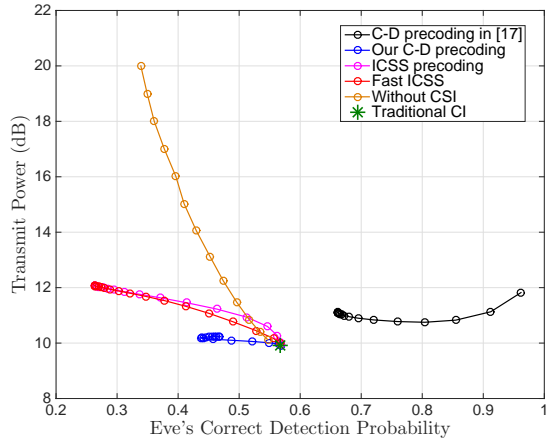}
	\caption{ Performance comparison between the schemes with and without eavesdropper's CSI for QPSK modulation, where $\rho=0.3$.}
	\label{Tradeoff_QPSK_noCSI}
\end{figure}

Finally, we investigate the scenario where the eavesdropper's CSI is unavailable. In Fig.~\ref{Tradeoff_QPSK_noCSI}, besides the  scheme proposed in Section \ref{sec_noCSI}, which is labeled ``Without  CSI'', we also reintroduce the  aforementioned CSI-dependent algorithms as baselines.  Another  baseline scheme, referred to as ``Traditional CI'', is also introduced. The  Traditional CI scheme aims at solving  optimization problem \eqref{equation-16} without constraints \eqref{eq-16-constraint-3} and \eqref{eq-16-constraint-4}. Thus, it only focuses on the  legitimate users' communication reliability without taking secrecy into consideration. First of all, we can observe that the secrecy performance of the  C-D precoding in~\cite{kat1} is even worse than the Traditional CI  scheme, which indicates that an inappropriate design of the security-aware precoder will  expose the transmitted symbol to the smart eavesdropper.  
Moreover, except for the C-D precoding in~\cite{kat1}, the other three CSI-dependent methods tend to converge  to the Traditional CI method as the transmit power decreases.  For the algorithm that does not use eavesdropper's CSI,  the minimum required power is always larger than that of the Traditional CI approach. This is because  additional power is needed to satisfy constraint \eqref{eq-35-constraint-1}, which in turn helps reduce the  eavesdropper's  detection probability. One can also observe that for the Without CSI scheme, enhancing transmit power is always beneficial for improving the secrecy. However, due to the absence of eavesdropper's CSI, more power is generally needed to achieve the same secrecy level as the two ICSS algorithms.  One exception  is when the eavesdropper's correct  detection probability is  larger than 0.5, where the Without  CSI scheme is more energy-efficient than at least one of the two ICSS algorithms.

\begin{figure}
	\centering
	\includegraphics[width=3.3in]{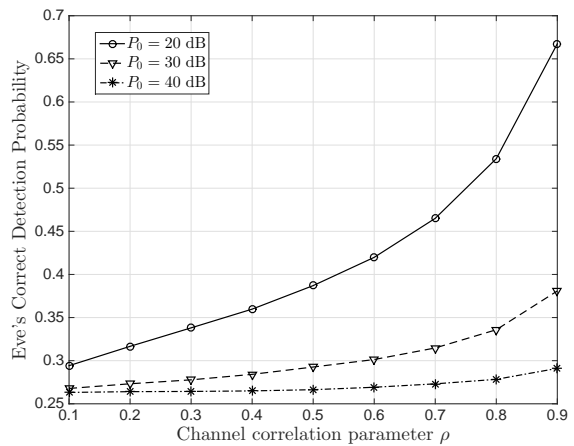}
	\caption{ Secure precoding without eavesdropper's CSI: Eavesdropper's correct detection probability versus channel correlation parameter for different $P_0$.}
	\label{rho_QPSK_noCSI}
\end{figure}

Fig.~\ref{rho_QPSK_noCSI} shows the secrecy performance of the Without  CSI approach versus the channel correlation parameter for different $P_0$. From Fig.~\ref{rho_QPSK_noCSI}  we can see that the eavesdropper's detection probability increases as $\rho$ increases, which is expected according to \eqref{equation-33}. Not only is the eavesdropper's received signal is more likely to be concentrated at the  symbol of interest as $\rho$ increases,   the 
power of the random term in \eqref{equation-33}, which acts as noise for blocking the eavesdropper, reduces as $\rho$ increases. Meanwhile, one can also observe that  improving $P_0$ is beneficial for reducing the eavesdropper's detection probability, which is consistent with the result in Fig.~\ref{Tradeoff_QPSK_noCSI} that  the  eavesdropper's detection probability decreases monotonically with an increases in the transmit power for the Without  CSI algorithm.

\section{Conclusions} \label{sec_conclusions}

In this paper, we restudied the CI-based secure precoding problem. Unlike the existing research, we considered a smart eavesdropper who can utilize statistical information for ML symbol detection. We first modified the existing CI-based precoding algorithm for a better utilization of the destructive interference. Then, we pointed out the security risk that both the original and our modified precoders have when faced with a smart eavesdropper. To combat the smart eavesdropper, a general  principle for designing security-aware precoders  was given. Two specific precoding  algorithms were then provided. In addition, the scenario without eavesdropper's CSI was also studied. Finally, we used  numerical results  to show the importance of the proposed schemes for mitigating a smart eavesdropper.



\begin{thebibliography}{10}
\providecommand{\url}[1]{#1}
\csname url@samestyle\endcsname
\providecommand{\newblock}{\relax}
\providecommand{\bibinfo}[2]{#2}
\providecommand{\BIBentrySTDinterwordspacing}{\spaceskip=0pt\relax}
\providecommand{\BIBentryALTinterwordstretchfactor}{4}
\providecommand{\BIBentryALTinterwordspacing}{\spaceskip=\fontdimen2\font plus
\BIBentryALTinterwordstretchfactor\fontdimen3\font minus
  \fontdimen4\font\relax}
\providecommand{\BIBforeignlanguage}[2]{{%
\expandafter\ifx\csname l@#1\endcsname\relax
\typeout{** WARNING: IEEEtran.bst: No hyphenation pattern has been}%
\typeout{** loaded for the language `#1'. Using the pattern for}%
\typeout{** the default language instead.}%
\else
\language=\csname l@#1\endcsname
\fi
#2}}
\providecommand{\BIBdecl}{\relax}
\BIBdecl

\bibitem{Spencer}
Q.~H. {Spencer}, A.~L. {Swindlehurst}, and M.~{Haardt}, ``Zero-forcing methods
  for downlink spatial multiplexing in multiuser {MIMO} channels,'' \emph{IEEE
  Trans. Signal Process.}, vol.~52, no.~2, pp. 461--471, Feb. 2004.

\bibitem{Sadek}
M.~{Sadek}, A.~{Tarighat}, and A.~H. {Sayed}, ``A leakage-based precoding
  scheme for downlink multi-user {MIMO} channels,'' \emph{IEEE Trans. Wireless
  Commun.}, vol.~6, no.~5, pp. 1711--1721, May 2007.

\bibitem{Christensen}
S.~S. {Christensen}, R.~{Agarwal}, E.~D. {Carvalho}, and J.~M. {Cioffi},
  ``Weighted sum-rate maximization using weighted {MMSE} for {MIMO-BC}
  beamforming design,'' \emph{IEEE Trans. Wireless Commun.}, vol.~7, no.~12,
  pp. 4792--4799, Dec. 2008.

\bibitem{SchubertBoche2004}
M.~{Schubert} and H.~{Boche}, ``Solution of the multiuser downlink beamforming
  problem with individual {SINR} constraints,'' \emph{IEEE Trans. Veh.
  Technol.}, vol.~53, no.~1, pp. 18--28, Jan. 2004.

\bibitem{Sidiro}
N.~D. {Sidiropoulos}, T.~N. {Davidson}, and {Zhi-Quan Luo}, ``Transmit
  beamforming for physical-layer multicasting,'' \emph{IEEE Trans. Signal
  Process.}, vol.~54, no.~6, pp. 2239--2251, Jun. 2006.

\bibitem{MasourosRat}
C.~{Masouros}, T.~{Ratnarajah}, M.~{Sellathurai}, C.~B. {Papadias}, and A.~K.
  {Shukla}, ``Known interference in the cellular downlink: {A} performance
  limiting factor or a source of green signal power?'' \emph{IEEE Commun.
  Mag.}, vol.~51, no.~10, pp. 162--171, Oct. 2013.

\bibitem{Masouros2}
C.~Masouros and G.~Zheng, ``Exploiting known interference as green signal power
  for downlink beamforming optimization,'' \emph{IEEE Trans. Signal Process.},
  vol.~63, no.~14, pp. 3628--3640, July 2015.

\bibitem{LiMasouros2018}
A.~{Li} and C.~{Masouros}, ``Interference exploitation precoding made
  practical: Optimal closed-form solutions for {PSK} modulations,'' \emph{IEEE
  Trans. Wireless Commun.}, vol.~17, no.~11, pp. 7661--7676, Nov. 2018.

\bibitem{Haqiqat}
A.~{Haqiqatnejad}, F.~{Kayhan}, and B.~{Ottersten}, ``Symbol-level precoding
  design based on distance preserving constructive interference regions,''
  \emph{IEEE Trans. Signal Process.}, vol.~66, no.~22, pp. 5817--5832, Nov.
  2018.

\bibitem{HJedda}
H.~Jedda, A.~Mezghani, A.~L. Swindlehurst, and J.~A. Nossek, ``Quantized
  constant envelope precoding with {PSK} and {QAM} signaling,'' \emph{IEEE
  Trans. Wireless Commun.}, vol.~17, no.~12, pp. 8022--8034, Dec. 2018.

\bibitem{Ang2}
A.~{Li}, C.~{Masouros}, F.~{Liu}, and A.~L. {Swindlehurst}, ``Massive {MIMO}
  1-bit {DAC} transmission: A low-complexity symbol scaling approach,''
  \emph{IEEE Trans. Wireless Commun.}, vol.~17, no.~11, pp. 7559--7575, Nov.
  2018.

\bibitem{Wyner}
A.~D. Wyner, ``The wire-tap channel,'' \emph{Bell Syst. Tech. J.}, vol.~54,
  no.~8, pp. 1355--1387, Oct. 1975.

\bibitem{Goel}
S.~Goel and R.~Negi, ``Guaranteeing secrecy using artificial noise,''
  \emph{IEEE Trans. Wireless Commun.}, vol.~7, no.~6, pp. 2180--2189, Jun.
  2008.

\bibitem{Wei}
W.~C. Liao, T.~H. Chang, W.~K. Ma, and C.~Y. Chi, ``{QoS}-based transmit
  beamforming in the presence of eavesdroppers: An optimized
  artificial-noise-aided approach,'' \emph{IEEE Trans. Signal Process.},
  vol.~59, no.~3, pp. 1202--1216, Mar. 2011.

\bibitem{J.Zhuu}
J.~Zhu, R.~Schober, and V.~K. Bhargava, ``Linear precoding of data and
  artificial noise in secure massive {MIMO} systems,'' \emph{IEEE Trans.
  Wireless Commun.}, vol.~15, no.~3, pp. 2245--2261, Mar. 2016.

\bibitem{ZhiDing}
S.~{Bashar}, Z.~{Ding}, and C.~{Xiao}, ``On secrecy rate analysis of {MIMO}
  wiretap channels driven by finite-alphabet input,'' \emph{IEEE Trans.
  Commun.}, vol.~60, no.~12, pp. 3816--3825, Dec. 2012.

\bibitem{kat1}
M.~R.~A. {Khandaker}, C.~{Masouros}, and K.~{Wong}, ``Constructive interference
  based secure precoding: {A} new dimension in physical layer security,''
  \emph{IEEE Trans. Inf. Forensics Security}, vol.~13, no.~9, pp. 2256--2268,
  Sep. 2018.

\bibitem{kat2}
M.~R.~A. {Khandaker}, C.~{Masouros}, K.~{Wong}, and S.~{Timotheou}, ``Secure
  {SWIPT} by exploiting constructive interference and artificial noise,''
  \emph{IEEE Trans. Commun.}, vol.~67, no.~2, pp. 1326--1340, Feb. 2019.

\bibitem{Ont}
A.~{Kalantari}, M.~{Soltanalian}, S.~{Maleki}, S.~{Chatzinotas}, and
  B.~{Ottersten}, ``Directional modulation via symbol-level precoding: A way to
  enhance security,'' \emph{IEEE J. Sel. Topics Signal Process.}, vol.~10,
  no.~8, pp. 1478--1493, Dec. 2016.

\bibitem{datong}
D.~{Xu}, P.~{Ren}, and H.~{Lin}, ``Combat hybrid eavesdropping in power-domain
  {NOMA}: {Joint} design of timing channel and symbol transformation,''
  \emph{IEEE Transactions on Vehicular Technology}, vol.~67, no.~6, pp.
  4998--5012, Jun. 2018.

\bibitem{CL}
J.~Zhang, B.~He, T.~Q. Duong, and R.~Woods, ``On the key generation from
  correlated wireless channels,'' \emph{IEEE Commun. Lett.}, vol.~21, no.~4,
  pp. 961--964, Apr. 2017.

\bibitem{bigM2}
D.~W.~K. {Ng}, Y.~{Wu}, and R.~{Schober}, ``Power efficient resource allocation
  for full-duplex radio distributed antenna networks,'' \emph{IEEE Trans.
  Wireless Commun.}, vol.~15, no.~4, pp. 2896--2911, Apr. 2016.

\bibitem{S.Boyd}
S.~Boyd and L.~Vandenberghe, \emph{Convex Optimization}.\hskip 1em plus 0.5em
  minus 0.4em\relax Cambridge, UK: Cambridge Univ. Press, 2004.

\bibitem{Jiang}
X.~{Jiang}, W.~{Zeng}, A.~{Yasotharan}, H.~C. {So}, and T.~{Kirubarajan},
  ``Quadratically constrained minimum dispersion beamforming via gradient
  projection,'' \emph{IEEE Trans. Signal Process.}, vol.~63, no.~1, pp.
  192--205, Jan. 2015.

\bibitem{complexity}
K.~{Wang}, A.~M. {So}, T.~{Chang}, W.~{Ma}, and C.~{Chi}, ``Outage constrained
  robust transmit optimization for multiuser {MISO} downlinks: Tractable
  approximations by conic optimization,'' \emph{IEEE Trans. Signal Process.},
  vol.~62, no.~21, pp. 5690--5705, Nov. 2014.

\bibitem{Zhou}
X.~{Zhou} and M.~R. {McKay}, ``Secure transmission with artificial noise over
  fading channels: Achievable rate and optimal power allocation,'' \emph{IEEE
  Trans. Veh. Technol.}, vol.~59, no.~8, pp. 3831--3842, Oct. 2010.

\bibitem{SCP1}
B.~R. Marks and G.~P. Wright, ``A general inner approximation algorithmfor
  nonconvex mathematical programs,'' \emph{Oper. Res.}, vol.~26, no.~4, pp.
  681--683, 1978.

\bibitem{SCP2}
Y.~{Cheng} and M.~{Pesavento}, ``Joint optimization of source power allocation
  and distributed relay beamforming in multiuser peer-to-peer relay networks,''
  \emph{IEEE Trans. Signal Process.}, vol.~60, no.~6, pp. 2962--2973, Jun.
  2012.

\end{thebibliography}
\end{document}